\documentclass[sigconf,screen]{acmart}
\acmConference[ESEC/FSE 2023]{The 31st ACM Joint European Software Engineering Conference and Symposium on the Foundations of Software Engineering}{11 - 17 November, 2023}{San Francisco, USA}

\usepackage{multirow}

\usepackage{tikz}
\usepackage{xcolor}

\usepackage{graphicx}
\usepackage{subfigure}

\usepackage{makecell}

\newcolumntype{D}{ >{\arraybackslash} m{4cm} }
\newcolumntype{C}{ >{\centering\arraybackslash} m{0.5cm} }
\usepackage{colortbl}

\usepackage{colortbl}

\definecolor{cellorange}{rgb}{ 1,  .949,  .8}
\definecolor{cellgreen}{rgb}{ .776,  .878,  .706}
\definecolor{cellblue}{rgb}{ .608,  .761,  .902}

\usepackage{ifthen}
\usepackage{xcolor}

\usepackage{fontawesome}

\usepackage{widetext}

\usepackage[titletoc]{appendix}

 \renewcommand\footnotetextcopyrightpermission[1]{} 
\begin{document}


\title{Software Architecture in Practice: Challenges and Opportunities}

\author{Zhiyuan Wan}
\orcid{0000-0001-7657-6653}
\affiliation{\institution{Zhejiang University}
  \city{Hangzhou}
  \country{China}
}
\email{wanzhiyuan@zju.edu.cn}

\author{Yun Zhang}
\authornote{Corresponding author.}
\affiliation{\institution{Hangzhou City University}
 \city{Hangzhou}
 \country{China}}
 \email{yunzhang@hzcu.edu.cn}

\author{Xin Xia}
\affiliation{\institution{Huawei}
  \city{Hangzhou}
  \country{China}}
\email{xin.xia@acm.org}

\author{Yi Jiang}
\affiliation{\institution{Huawei}
  \city{Shanghai}
  \country{China}}
\email{jiangyi54@huawei.com}

\author{David Lo}
\affiliation{\institution{Singapore Management University}
  \city{Singapore}
  \country{Singapore}
  }
 \email{davidlo@smu.edu.sg}
\begin{abstract}
  Software architecture has been an active research field for nearly four decades, in which previous studies make significant progress such as creating methods and techniques and building tools to support software architecture practice. Despite past efforts, we have little understanding of how practitioners perform software architecture related activities, and what challenges they face. Through interviews with 32 practitioners from 21 organizations across three continents, we identified challenges that practitioners face in software architecture practice during software development and maintenance. We reported on common software architecture activities at software requirements, design, construction and testing, and maintenance stages, as well as corresponding challenges. Our study uncovers that most of these challenges center around management, documentation, tooling and process, and collects recommendations to address these challenges.
\end{abstract}

\maketitle

\section{Introduction}\label{sec:introduction}
Software architecture refers to a collection of design decisions that affect the structure, behavior, and overall quality of a software system~\cite{rebouccas2017does,clements2003documenting}, serving as the foundation for subsequent decisions~\cite{plosch2014value}. 
The research field of software architecture has achieved tremendous progress since its inception in the 1980s~\cite{shaw2006golden}. In the 2000s, there has been a paradigm shift in understanding the essence of software architecture, moving from a purely technical view to a socio-technical view.
The early paradigm centered around the tangible outcomes of software architecture practice, such as the structure and behavior of software systems, their components and connectors, and the use of views, architecture description languages, design methods and patterns~\cite{capilla2016ten}.
The subsequent paradigm concerned how stakeholders reach the outcomes through software architecture practice, specifically how they reason about their choices and make architectural decisions~\cite{jansen2009enriching}.
Previous studies have put effort into developing methods and tools for architecture representation and documentation~\cite{jansen2009enriching,cai2018design}, analysis and evaluation~\cite{patidar2015survey,kazman2012scaling}, recovery and optimization~\cite{aleti2012software,garcia2013comparative,lutellier2015comparing}, as well as knowledge management and decision making~\cite{capilla2016ten,van2016decision} to support software architecture practice.

Despite these efforts, we have little understanding of how practitioners actually perform software architecture practice during the development and maintenance of software systems, and what challenges they encounter in practice. 
To systematically explore the challenges and point improvement towards better practice, we conducted interviews with 32 participants involved in the design, implementation and maintenance of software architectures.
Our research question is \emph{what are the software architecture related activities performed in practice, and the corresponding challenges faced by practitioners}?
Interview participants come from 21 organizations of varying sizes, from small startups to large technology companies. They have diverse roles in software development and maintenance, including architect, development, testing and project management.
During the interviews, we explored various aspects of software architecture practice, including the architectural styles applied, techniques and processes followed, and tools utilized. We also sought to identify where challenges arise in software architecture practice during software development and maintenance.

We observe that challenges in software architecture practice surface at different stages of software development and maintenance process: (1) Evolution and changes of software requirements at the requirements stage; (2) Design documentation, requirements decomposition, and architecture analysis and evaluation at the design stage; (3) Architecture conformance checking, continuous architecture monitoring, and code quality at the construction and testing stage; and (4) Architecture erosion and refactoring at the maintenance stage. Even though architectural styles and software architecture practice differ substantially across organizations, we find common patterns of architectural styles and associated challenges.
Overall, our observations suggest four themes that would benefit from more attention with respect to management (\faGroup), documentation (\faFileText), tooling (\faCogs), and process (\faCalendar): (1) Fostering an organization-wide culture of building high-quality architecture; (2) Paying more attention to the up-to-dateness and traceability of design documentation; (3) Adopting and developing effective tools to support software architecture practice; and (4) Improve the reuse of architecture knowledge to facilitate architecture design.

In summary, this paper makes the following contributions: (1) We identified challenges in software architecture practice during software development and maintenance through interviews with 32 practitioners, triangulated with a literature review; (2) We provided recommendations for improving software architecture practice and outlined future avenues of research. 

The remainder of the paper is structured as follows.
In Section \ref{sec:related}, we briefly review related work. In Section \ref{sec:methodology}, we describe the methodology of our study in detail. Sections \ref{sec:result_requirements}, \ref{sec:result_design}, \ref{sec:result_construction_testing} and \ref{sec:result_maintenance} present the results of our study. We discuss the implications of our results in Section \ref{sec:discussion} and  any threats to the validity of our findings in Section \ref{sec:threat}. Section \ref{sec:conclusion} draws conclusions
and outlines avenues for future work.
 
\section{Related Work}\label{sec:related}

\noindent{\textbf{Structural Construction of Software Systems.}}
Software architecture practice embraces the concept of architecture \emph{view}, which represents a partial aspect of the high-level structures of a software system~\cite{clements2003documenting}. Researchers propose various approaches for documenting the relevant views of software architecture to help practitioners successfully use it, maintain it, and build a software system from it, including unified modeling language, architecture description languages~(e.g., \cite{medvidovic2000classification}), and domain-specific languages~(e.g., \cite{membarth2015hipa}). Despite the development of formal approaches for specifying software architectures, in practice, natural language is widely used in documenting software architectures, sometimes accompanied by diagrams of informal models~\cite{muszynski2022study}. 
The benefits of software architecture documentation have been widely investigated in previous studies, e.g., serving as an educational tool to introduce new team members to a software project~\cite{clements2003documenting} and reduce entry barriers for new contributors in OSS development~\cite{kazman2015evaluating}. 
\emph{In this paper, we explore how software practitioners capture and document software architecture.}

\noindent{\textbf{Design Decisions.}}
The early literature has made great efforts to develop approaches and tools for capturing architecture design decisions explicitly, including formal models (e.g., \cite{jansen2005software}) and description templates~(e.g., \cite{tyree2005architecture}). The formal models and description templates have consensus on capturing rationale, constraints, and alternatives of architectural design decisions~\cite{shahin2009architectural}.
Recently, researchers propose approaches for the automatic extraction of design decisions from different information sources (e.g.,~\cite{li2020automatic}), e.g., email archives, issue management systems, commit messages, and Stack Overflow. 
In addition, researchers conduct empirical studies to investigate architectural decision-making practice including decision-making process~(e.g.,~\cite{groher2015study}), methods (e.g.,~\cite{drury2017examining}), and tools~(e.g.,~\cite{gaubatz2015automatic}), and shift their focus towards addressing the social and psychological aspects of architectural decision making behaviors including naturalistic and rational decision making~\cite{moe2012challenges}, cognitive biases~\cite{mohanani2014requirements}, and group decision making~\cite{borte2012role,rekhav2014study}.
\emph{In this paper, we explore how software practitioners make architectural design decisions.}

\noindent{\textbf{Software Evolution.}} 
From an evolution perspective, architecture analysis and evaluation is a preventive activity to improve qualitative attributes, delay architectural decay, limit the effect of software aging, and identify architectural drift and erosion of software systems~\cite{maccari2002experiences,tonu2006evaluating}. Previous studies propose a wide spectrum of approaches and tools for architecture analysis and evaluation, including extracting intuitions and experiences of stakeholders and further using their tacit knowledge (e.g.,~\cite{bouwers2010lightweight}), creating scenario profiles for a concrete description of quality attributes (e.g.,~\cite{olumofin2007holistic}), architecture conformance and compliance checking (e.g.,~\cite{terra2015recommendation}), and using various quality metrics (e.g., MoJoFM~\cite{wen2004effectiveness}).

Architectural changes occur during regular development and maintenance activities, which involve a wider spectrum of code components, as well as dependencies among them than local code changes~\cite{schmitt2020arcade}.
It is complex for practitioners to comprehend the scopes and impacts of architectural changes~\cite{tang2014software,uchoa2021predicting}. The complexity elevates the cost and effort of implementing architectural changes across the software development lifecycle~\cite{williams2014examination}. 
\emph{In this paper, we explore how software practitioners perceive software evolution, as well as how they conduct architectural changes and address the evolution problems from the architectural perspective.}

\noindent{\textbf{Architecture Knowledge Management.}}
It has been suggested that ``if it [an architecture design] is not written down, it does not exist''~\cite{clements2010documenting}, thus to prevent knowledge vaporization and architectural drift, a plethora of works on capturing different types of architecture knowledge have emerged for modeling design decisions and rationale. The uses for architecture knowledge span four broad categories, i.e., sharing, compliance, discovery and traceability~\cite{de2007architectural}.

A shared understanding of architecture knowledge among stakeholders of software systems alleviates miscommunication and information overload, especially for the software systems that are produced by practitioners from different geographic areas~\cite{dutoit2001knowledge}. Due to the large number of stakeholders and the dispersion of knowledge, it is challenging to effectively share and reuse architecture knowledge~\cite{miksovic2011architecturally}.
Architecture knowledge enables evaluation and review of architecture compliance (e.g.,~\cite{vogel2011software}), in terms of requirements missing and conflicting, conflicts between requirements and design, and violation of design principles.
The discovery of architecture knowledge allows reasoning and uncovering design problems and alternatives~\cite{capilla2016ten}.
Traceability of architecture knowledge improves understandability of software architectures, helps to locate relevant knowledge, enables impact analysis of architectural changes, and facilitates design review, evaluation and assessment~\cite{ramesh2001toward,li2013application}.
\emph{In this paper, we explore how software practitioners manage architecture knowledge.} 
\section{Methodology}\label{sec:methodology}
We adopted a qualitative research strategy to explore software architecture practice in software development and maintenance and corresponding challenges, with interviews of software practitioners in the industry.
Our study consists of four steps: 
(1) We prepared an interview guide informed by an initial literature review, (2) We conducted interviews, (3) We triangulated results with literature findings, and (4) We validated the findings with our interview participants. 
We based our research on Straussian Grounded Theory~\cite{strauss1994grounded,corbin1990basics}, which derives research questions from literature, analyzes interviews with open and axial coding, and consults literature throughout the process.
Specifically, we simultaneously conducted interviews and reviewed literature, utilizing immediate and continuous data analysis, making constant comparisons, and refining our codebook and interview guide throughout the study.

\noindent\textbf{Step 1: Scoping and Interview Guide.} To scope our research and prepare for the interviews, we searched for software architecture practice discussed in the existing literature on software engineering (Section \ref{sec:related}). 
In this phase, we selected 28 papers opportunistically through keyword search and our personal knowledge of the field. 
We applied standard open coding process~\cite{corbin1990basics} to identify sections in the papers that potentially relate to software architecture practice.
Although most papers did not directly address challenges in software architecture practice, we marked discussions that potentially related to challenges, e.g., inadequate knowledge management for design decisions.
We then analyzed and condensed these codes into the stages of software development and maintenance, and developed an initial codebook and interview guide (provided in Supplement A and Supplement B).
\noindent\textbf{Step 2: Interviews.} 
The first author conducted a series of interviews with 32 software practitioners from 21 organizations\footnote{The interviews were approved by the relevant institutional review board (IRB) Participants were instructed that we wanted their opinions; privacy and sensitive information would not be intentionally mentioned.}, 4 in-person and 28 online interviews. Each interview is 45 to 60 minutes long. The interviews were semi-structured and made use of the interview guide, which was sent to the participants before the interviews, in line with previous studies~\cite{wan2022motivates,wan2021smart,wan2019does}.
All the interview participants are involved in industrial software projects. The demographics of interview participants are summarized in Table \ref{tab:demographics}, with details found in Supplement C. Note that some participants take more than one role in their projects.

\noindent\textbf{Participant Selection.}
We selected participants with the maximum variation sampling method~\cite{suri2011purposeful} of purposeful sampling to cover participants with different job roles and in various types of organizations. 
We adapted our recruitment strategy throughout the study based on our findings in the interviews. In the later stages of the study, we focused on specific roles and organizations to fill gaps in our understanding.
Participant selection, data collection and data analysis continued until saturation was reached and a rich description of experience had been obtained. New codes did not appear anymore in data analysis; the set of codes was considered stable. We recruited potential participants with necessary software architecture knowledge through personal networks and recommendations from previous participants. 
We separately interviewed multiple participants within the same organization to get different perspectives. 
Given that some organizations with multiple business units may adopt different software practice, we recruited several participants from each business unit for those organizations.
For confidentiality, we refer to organizations by number, and participants by $PI_j$ where $I$ refers to the organization number and $j$ distinguishes participants from the same organization.

\noindent\textbf{Data Analysis.} 
All interviews were recorded, transcribed and analyzed using constant comparative method. Data collection and data analysis took place simultaneously. 
After each interview, the first author transcribed the recording of the interview, open coded the transcript using NVivo qualitative analysis software~\cite{qsr20231nvivo}. Specifically, the first author broke the transcript down line-by-line into concepts termed meaning units, labeled the units with codes, and continuously compared similar codes. To ensure the quality of codes, the second author reviewed the initial codes created by the first author and provided suggestions for improvement. 
These suggestions were discussed and incorporated into the codes.  
After the open coding stage, we generated a total of 272 unique codes of software architecture activities and challenges -- 15 to 79 codes for each interview. 
Next, we conceptualized the resulting codes by specifying the relationship between them and integrating them into categories, e.g., \emph{design documentation} and \emph{architecture erosion}. 
Finally, we mapped these categories to software development and maintenance phases, e.g., \emph{software requirements} and \emph{software construction and testing}.
In addition, we created visualizations of architectural styles in each organization (provided in Supplement D), and used these visualizations to explore whether the concepts associated with certain types of architectural styles. 

\noindent\textbf{Step 3: Triangulation with Literature.} 
Triangulation refers to the use of multiple methods or data sources in qualitative research to increase the credibility and validity of research findings~\cite{patton1999enhancing}, which is also used in grounded theory research on software engineering~(e.g., \cite{nahar2022collaboration}).
We used methodological triangulation, which involves the use of two methods to gather data, i.e., interviews and literature review. 
Specifically, as we gained insights from interviews, we returned to the literature to identify related discussions and potential solutions to triangulate our interview results. 
We pursued a best-effort approach that relied on keyword search for themes that surfaced in the interviews, as well as backward and forward snowballing. Consequently, we identified 65 papers as possibly relevant and coded them with the evolving codebook. 
The data from the papers were used to confirm and support the findings of the interviews, representing triangulation. 
The complete list of the 65 papers can be found in Supplement E.
\noindent\textbf{Step 4: Validity Check with Interview Participants.} For assessing fit and applicability as defined by Strauss and Corbin~\cite{corbin1990basics} and validating our findings, we returned to our participants after creating a full draft of this paper. We presented the participants with both the full draft and a summary of the challenges and recommendations that emerged during the interviews, along with questions that prompted them to look for correctness and areas of agreement or disagreement (i.e., fit), and any insights gained from reading experiences of the other companies, roles, and the overall findings (i.e., applicability). 
Specifically, participants were asked to indicate agreement or disagreement by placing a tick or cross next to each challenge or recommendation based on their realm of experience; they were also asked if they had any insights to add. All participants indicated general agreement and six responded with comments, several explicitly reaffirmed some findings. We incorporated three minor suggested changes to details in the recommendations.

\begin{table}[t]
  \centering
  \caption{Participant and company demographics.}
\small
    \begin{tabular}{p{10em}p{17em}}
\toprule
    \textbf{Type} & \textbf{Break-down} \\
    \midrule
    Participant Role (32) & Architect (17), Development (13), Project Management (11), Testing (2) \\
    Participant Seniority (32) & 10 years of experience or more (14), 5-10 years (17), under 5 years (1) \\
    Company Type (21) & Big tech (7), Non IT (6), Mid-size tech (5), Startup (3) \\
    Company Location (21) & Asia (14), North America (5), Europe (2) \\
    \bottomrule
    \end{tabular}\label{tab:demographics}\end{table}

\section{Software Requirements}\label{sec:result_requirements}

Some software requirements, particularly certain non-functional ones, have a global scope in that their satisfaction cannot be allocated to a discrete component in a software system. A requirement with a global scope could affect the software architecture and the design of many components.
\noindent\textbf{Unpredictable evolution and changes of software requirements complicate architecture design} (\faGroup, \faCalendar).
Architects perform architecture design towards a visionary future by foreseeing the potential evolution and changes in software requirements. 
Practitioners expect software systems to evolve and iterate spontaneously as requirements change over time (P1a, P2a, P3b, P3e, P4a, P4d, P6a, P7a, P10a, P12a, P13a, P14a, P15a, P17a, P18a).
For example, P1a (architect) shared, ``\emph{we had to think about scaling up the capacity, handling more users, and dealing with increased concurrency in our architecture design ... we were looking ahead, maybe in the next 2 to 3 years, considering that the business requirements would evolve.}''However, requirements sometimes tend to change and evolve in an unpredictable way, including the increasing variety of users (P1a, P10a, P11a), the change in business requirements of the whole industry (P4a, P11a, P14a), and the evolution of technology stacks (P3e, P7a, P8a, P9a, P11a, P12a, P13a, P15a). 
The unpredictable evolution and changes of requirements sometimes make it infeasible to deliver new features or meet the quality requirements with current architectures, as frequently mentioned in the literature~\cite{dasanayake2019impact,szlenk2012modelling,de2012controlling,mens2008software,tang2011software,arcelli2015control,baudry2014diversify}.

In practice, even experienced architects cannot design a perfect software architecture that can support evolution and changes of requirements in the long-term future, especially with modest time and resources (P1a, P2a, P3e, P4d, P8a, P11a, P12a, P13a, P14a, P15a). 
Some participants explained that the challenges arise because ``\emph{requirements violate the original assumptions about the expected quality attributes of the software systems, e.g., capacity, number of concurrent users, and TPS [number of transactions per second]}'' (P2a, developer).
Other participants explained, a ``perfect'' architecture could cost far more than affordable. Given the development of technology stacks are fast and evitable,  aggressive adoption of the latest technology increases development cost (P6a, P11a).
For instance, transaction-based features can be hardly implemented with a non-transaction-based architecture of a software system (P3e). 

\noindent\textbf{Recommendations.} 
It is important for architects to make tradeoffs between multiple factors with awareness of requirements volatility and unpredictability (\faCalendar). 
It seems beneficial to adopt more formal architecture documentation for capturing the tradeoffs in architecture design, and constructing trace links between requirements and design decisions (\faFileText), which has been suggested in the literature~\cite{fischer2001knowledge,babar2007tool,bonnema2014communication}. The explicit capturing of tradeoffs and their rationales facilitates the communication of essential design decisions among various stakeholders, and the reuse of architecture knowledge. 
A practical strategy could be designing a software architecture that can support the evolution and changes in requirements for one to three years in the future (P2a).
Some participants suggest a reactive strategy to overcome requirements volatility and unpredictability -- a standard process for adapting, refactoring, and retiring software architectures (\faCalendar) (P2a, P3g).  
 \section{Software Design}\label{sec:result_design}
Jan Bosch emphasized that ``designing a system can be viewed as a decision process''~\cite{bosch2004software}. Practitioners work with various stakeholders to make design decisions with a high-level view of both business and technical aspects of software systems. We observed challenges in design documentation, design principle application, and design quality analysis and evaluation with respect to software architectures.
\subsection{Design Documentation}\label{sec:result_design_doc}
Design documentation describes information about the design, interfaces and functionality of software to support practitioners in their development and maintenance activities, by which the
diverse stakeholders communicate with the design team and with each other~\cite{selic2009agile}. 

\noindent\textbf{Use of models and tools is inadequate to ensure the completeness of architecture documentation} (\faFileText, \faCogs).
Most organizations provide unified templates for architectural specifications, but no adequate support of models and tools to ensure the completeness of architecture documentation. 
As reported by participants, the common parts of architectural specifications across organizations include views of software architecture, interface specifications, and rationales behind design decisions. 
Some large-scale organizations allow customization to organizational-wide templates for different business units (e.g., P3d (architect): ``\emph{we've tweaked the company's templates to better suit our specific business needs.}''). 
Nonetheless, no participants reported that they use any formal models or tools to make sure that architecture documentation is complete for communication.
Some participants mentioned that sometimes they cannot find the relevant information in architecture documentation (P9a, P19b). The incompleteness of design documentation is also discussed in the literature~(e.g.,~\cite{zhi2015cost}).

\noindent\textbf{Architecture documentation becomes obsolete as software evolves} (\faFileText,~\faCalendar).
Participants reported that design documentation tends to suffer from up-to-dateness problems. Most participants observe the inconsistency between design documentation and code implementation as software systems evolve, e.g., when integrating new features into the systems and bug fixing (P3b, P3e-g, P4c, P6a-b, P7a, P8a, P9a, P10a, P11a, P12a, P14a, P18a, P19a-b, P20a, P21a). The documentation-code inconsistency would confuse the developers who perform development tasks.
Moreover, a few participants in agile projects reported that documentation tends to be missing for new features or components due to a fast development pace. P6a (developer) explained, ``\emph{the traditional modeling method like UML becomes a significant roadblock to fast-paced design in agile development}''.
Whereas the literature discussed challenges in providing up-to-date software documentation (e.g.,~\cite{aghajani2019software}), our interviewees were concerned about  the inconsistency between design documentation and code implementation in particular.

\noindent\textbf{Inadequate tool support for sharing, version control, and tracing of scattered design documentation} (\faFileText, \faCogs).
Design documentation is written and organized with a wide variety of methods across and within organizations as reported by our participants (P3b, P3e, P3g, P4a, P5a, P6a, P7a, P11a, P13a, P14a). We observe that most participants use traditional text processing software to write design documentation, and version control systems to keep track of changes in documents. 
For example, P3g (project manager, architect) described, ``\emph{we simply use Microsoft Office Word to capture our design documents}.'' 
The ineffective methods for organizing design documentation further impact its usability for readers, e.g., information findability and content searchability, as well as its usefulness for practical use in software development and maintenance.  These findings align with literature past observations that design documents are typically written in natural languages with supporting diagrams~\cite{ding2014knowledge}.
In most organizations, practitioners performed detailed design for a module or microservice with the architecture documentation of interface-level design as input. Some of the practitioners reported they included detailed design documentation in code comments (P2a, P6a), but few intentionally build the trace links between architecture and detailed design documentation.

\noindent\textbf{Recommendations.} 
Architecture documentation quality is important to understand and evolve software architectures, as well as training and education of developers (\faFileText). Participants emphasized the completeness, up-to-dateness, usability and usefulness of design documentation in our interviews (P2a, P3a-g, P4a, P5a, P6a, P11a, P12a, P19a). 
The importance of high-quality documentation is frequently discussed in the literature (e.g., \cite{clements2003documenting}).  As for layered architectures, interface specifications provide the standardized mechanism in which subsystems can effectively communicate with each other and enable them to operate as independent modules (P3a, P3f).
When it comes to microservices architectural style, some organizations limit the scope of interface specifications at the microservice level in a software system (P4a, P5a).

Some organizations have a standard process to make design documentation up to date, in which developers cannot merge their code unless the corresponding design documentation has been updated, and use wikis to host design documentation  (\faCalendar) (P4c, P19b). 
Placing documentation in standard locations is an effective practice to help practitioners locate it, as suggested in the literature~(e.g., \cite{aghajani2019software}).
Some organizations adopt collaborative writing tools for the generation and sharing of design documentation (P4d, P6a-b, P11a, P14a, P19a-b;~\faCogs). In contrast, a few participants suggest applying the ``code as documentation'' principle\footnote{\url{https://martinfowler.com/bliki/CodeAsDocumentation.html}} to avoid extra cost and efforts for writing and maintaining design documentation (P4a, P5a;~\faCalendar). 
\subsection{Design Principles} \label{sec:result_design_principles}
Practitioners applied a wide range of design principles when designing software architectures, including abstraction, coupling and cohesion, decomposition and modularization, encapsulation and information hiding, separation of interface and implementation, and separation of concerns.

\noindent\textbf{Unclear boundaries between architectural elements in software systems} (\faGroup,~\faCalendar).
Some practitioners observe challenges in understanding underlying businesses when it comes to the decomposition of architectural components (P3a, P4d, P5a, P8a, P9a, P11a, P19a, P21a). As P3a (architect) explained, ``\emph{one cannot elegantly decompose a software system by simply collecting and listing a bunch of business scenarios}.'' The misunderstanding of businesses leads to unclear boundaries between architectural components in a software system, and further affects the quality and productivity of the detailed design of the system. 
Software decomposition has been extensively studied in prior research of software engineering~\cite{lung2004applications,mitchell2006automatic}, 
in which software clustering has become an active research area. software clustering is defined as the process of decomposing large software systems into smaller, manageable, highly cohesive, loosely coupled, and feature-oriented subsystems~\cite{cui2011applying}. 
A recent study~\cite{sarhan2020software} presents a systematic literature analysis to structure and categorize the state-of-the-art research evidence of software clustering over the past decade.

As for a microservice architecture, it is challenging to decompose a software system into the optimal number of microservices with appropriate levels of granularity (P3a, P4d, P5a, P11a)~\cite{li2019dataflow,chen2017monolith}. On the one hand, the granularity and number of microservices would affect the code quality and cost efforts in subsequent development activities. P5a (architect) shared, ``\emph{bigger microservices tend to have tedious interfaces, while smaller microservices tend to introduce more cost and efforts when testing and addressing issues when integrating microservices}.'' 
On the other hand, the granularity and number of microservices could make an impact on the satisfaction levels of non-functional requirements, e.g., the latency of service invocation.

\noindent\textbf{Interdisciplinary knowledge is required to lower coupling and improve cohesion of software} (\faGroup,~\faCalendar).
A common theme in the interviews is that it is challenging to design a software system that is loosely coupled and highly cohesive (P2a, P3a-b, P3f-g, P4a-b, P6a, P11a).
As participant P3a (architect) explained, ``\emph{business requirements change over time in different frequencies ... a component [in a software system] tend to be highly coupled with others [as the system evolves] if it is responsible for both frequently and rarely changed requirements.}''
A highly coupled and low-cohesive software architecture could further lead to difficulty in the planning and management of subsequent software development activities, like detailed design and implementation.
Identifying the business requirements with potential changes in the future requires interdisciplinary knowledge of both architecture design and businesses for practitioners.

\noindent\textbf{Recommendations.} 
Decomposition and modularization are important to architecture design as a principle most frequently mentioned in our interviews (\faCalendar). 
It is crucial to place different functionalities and responsibilities in different components of software systems (P3a-d, P3f-g, P4a, P5a).
The decomposition of software systems occurs from two perspectives, vertically decomposing a software system into layers and horizontally decomposing the system into smaller components. 
For software systems with layered architecture styles, practitioners usually emphasize vertical decomposition for isolating software from physical hardware changes. The isolation further makes it easier to move the software components between different hardware solutions for a variety of application domains. 
Conversely, for cloud-based and microservice systems, practitioners emphasize the horizontal decomposition of systems into microservices and components with a deep understanding of the underlying business requirements of the systems (P5a;~\faCalendar). Practitioners should make tradeoffs of balancing the granularity and number of microservices, as well as balancing the non-functional requirements for individual microservices and the satisfaction level for the overall system~\cite{hassan2016microservices,chen2017monolith}.
To identify and tailor services based on business requirements, some organizations use Domain-Driven Design (DDD) as a flexible methodology to create a high
level of microservices architecture design with an iterative design and development process (P3a, P3f, P4a, P5a; \faCalendar). 

Coupling and cohesion are also frequently discussed as important design principles in our interviews (\faCalendar). For different architecture styles, coupling measures interdependence among architectural elements of different levels of granularity, e.g., components, modules, layers and microservices, while cohesion measures the strength of association of architectural elements within a particular scope (P2a, P3a-b, P3f-g, P4a-b, P6a). For instance, several organizations enforce the coupling and cohesion principle in terms of separate compilation, testing, release and deployment of modules (P3g). 
Participants applied a variety of strategies to lower coupling and improve the cohesion of architectural elements. Some participants suggested designing independent architectural elements for change-prone businesses when decomposing a software system (P3a, P3g). 
To improve the cohesion of a software system, some participants suggested designing a module for common functionalities across businesses, which interacts with other modules or microservices (P3b, P4b, P6a).

\subsection{Design Quality Analysis and Evaluation} 
Practitioners take into account various quality attributes that contribute to the quality of software architecture, including quality attributes at runtime and those not discernible at runtime. 
To effectively analyze and evaluate the quality of software architectures, some organizations adopt automated techniques and tools, as well as measures for quantitative estimation.

\noindent\textbf{Architecture review requires a standard process, active involvement of external experts, and tool support} (\faCalendar,~\faGroup,~\faCogs).
The majority of our participants reported that architects and design teams perform architecture reviews in their projects, yet few mentioned the involvement of external consultants in architecture reviews. 
Architecture review and evaluation intends to uncover risks and issues in software architectures before they cause tremendous costs later in the software engineering life cycle. 
Large-scale organizations tend to conduct formal sessions of architecture review within development teams on a regular basis (P3a, P4c-d, P11a, P19a), but smaller organizations set up informal architecture review irregularly, e.g., once after the completion of architecture documentation and before coding. For instance, P1a (architect) shared, ``\emph{informal architecture reviews would take place once every one to two years during software maintenance, planning for the evolution of software architectures.}''

Participants rely heavily on experience in architecture review and evaluation process, manually inspecting requirements, multiple potential solutions, and rationale behind architectural design decisions in architecture documentation~(P1a, P2a, P3a, P3d, P4b, P6a, P9a, P16a).
For example, P3d (architect) stated, ``\emph{senior engineers and architects often leverage their expertise to evaluate and critique architectures, and ultimately drive architectural design choices ... their experience guides them in assessing and shaping the overall architectural landscape.}''
The widespread use of experience-based reasoning aligns with findings in the literature~(e.g.,~\cite{babar2009software}). 

\noindent\textbf{Lack of effective and apply-to-all quantitative measures} (\faCogs, \faCalendar).
Literature has proposed various metrics to measure the quality of software architectures, including coupling~\cite{mo2016decoupling} and 
cohesion~\cite{perepletchikov2007cohesion}.
Most participants perceive the importance of measures to effectively quantify various aspects of software architectures, but identify two challenges in applying architecture measures in practice (P1a, P3a, P3b, P3d-g, P4a-b, P11a, P14a, P19a). 
First, it is challenging to propose quantitative measures that accurately reflect the in-depth problems in software architectures (P3a, P4a-b). In some organizations, the implemented quantitative measures for architecture quality tend to be superficial and lack of theoretical basis. The challenges come from complexity in the understanding of business logic as well as the architectures of software systems.
Second, no general criteria exist concerning good-quality architectures across software systems, making it challenging to enforce organizational-wide measures for architecture evaluation (P3a, P4a-b). Organizations do not enforce standard architectural design across business units, leading to a variety of architectural styles within organizations. P3a (architect) explained ``\emph{the reasonable decomposition of modules is closely dependent on the underlying scenarios ... the reasonable decomposition of microservices is also related to businesses ... No measures apply to all}.'' In some organizations that enforce organizational-wide measures, participants would adopt a whitelist approach to exclude special cases whenever the measures do not apply. 

\noindent\textbf{Recommendations.} 
Practitioners need to consider both runtime and non-runtime quality attributes of software systems, but with different priorities in architecture review (\faCalendar).  
Performance is the runtime quality attribute most frequently mentioned in the interviews (P1a, P2a, P3a, P3d, P3g, P4a, P5a, P6a, P11a, P15a, P19b), with example terms of latency and transactions per second. 
Other runtime quality attributes are also discussed, e.g., functionality (P3d, P3g, P4a, P4c, P5a, P11a), availability (P1a, P3a, P3d, P5a, P4c, P11a), elasticity (P5a, P11a), and security (P1a, P13a).
Meanwhile, most participants emphasized the importance of extensibility and maintainability as the non-runtime quality attributes of software systems (P1a, P3a, P3d, P3f-g, P4a, P6a). Some participants also considered reusability (P2a, P3f-g), compatibility (P3e), and testability (P3g).
It is also recommended to involve external experts from other teams to review and evaluate the software architectures, and provide prioritized recommendations for architecture improvement (\faGroup,~\faCalendar).
For instance, P2a (developer) suggested that ``\emph{the design teams should collect feedback about API design and encapsulation from the downstream users}.''

In terms of tool adoption to support architecture review and measurement, some participants suggested using simulation and prototyping techniques to evaluate potential solutions to support architecture decisions (P3a, P4a-c, P12a). A few participants suggested using static analysis tools for visualizing architectures (P1a), and checklists when conducting architecture reviews (P5a, P13a). 
Given thresholds for measures of architecture quality may differ across software systems, some participants suggest only applying a limited set of measures in automated tools for estimating fundamental architecture quality within organizations, yet leave the left as a reference for practitioners to make decisions (P3e, P3f). Other participants suggest using whitelists to exclude special cases from the automatic measurement of architecture quality (P3b).

 \section{Software Construction and Testing}\label{sec:result_construction_testing}
The software construction and testing phase relies on the outcomes of the software design process. We found many challenges during this phase stemming from architecture conformance checking, architecture monitoring, and construction quality. 

\subsection{Architecture Conformance Checking}\label{sec:result_construction_testing:conformance}
Architecture conformance checking aims to ensure the consistency between the implemented architecture of a software system and its intended architecture as the system is implemented. 
The diverges of implemented architecture from the corresponding intended architecture could lead to software architecture erosion~\cite{perry1992foundations}. 

\noindent\textbf{Automated architecture conformance checking is rare} (\faCogs).
Despite the common awareness of architecture divergence, participants rarely rely on automated tools to check architecture conformance in the software construction phase (P1a, P3b-c, P3e, P3f-g, P4a-c, P6b, P11a, P12a, P13a, P15a, P19b). 
Instead, some organizations rely heavily on periodical manual inspection, in line with the observations of previous studies (e.g., \cite{caracciolo2014software}). In the manual inspection for architecture conformance, inspectors rely on the detailed description of code changes, especially the architectural changes, thus the code contributors should provide comprehensive information for their changes. P4b (architect) explained, ``\emph{the detailed description [of architecture-level code changes] would save the time of architects to evaluate whether the code changes conform with the intended architecture}.''

\noindent\textbf{Obsolete documentation and lack of traceability hinder automation of architecture conformance checking} (\faFileText, \faCalendar).
Some participants report several potential obstacles that hinder the adoption of automated tools (P3b, P3g, P4a, P4c, P19a). First, the obstacles come from the up-to-dateness of architecture documentation and specifications. P4a (project manager, architect) explained, ``\emph{developers tend to forget updating architecture documentation when evolving software architectures because of deadline pressures}.'' The obsolete architecture documentation and specifications are unreliable for automated architecture conformance checking.
Second, no traceability links between artifacts exist to support locating the artifacts that contribute to the resulting architectural inconsistency. For instance, P3b (architect)  and P3g (project manager, architect)  illustrated, ``\emph{no standard process or tool support to build trace links between design decisions and their implementation}.''

\noindent\textbf{Recommendations.} 
Knowledge vaporization could cause inconsistency between the intended architecture and implemented system (\faGroup). Some participants attribute knowledge vaporization to high developer turnover in software development projects (P3e, P4a-b, P6b, P12a, P13a, P19a). Developer turnover aggravates architecture divergence by losing knowledge about project contexts including system requirements and architectural decisions~\cite{strasser2014mastering}. A poor understanding of project contexts also hinders knowledge transfer among team members.

\subsection{Continuous Architecture Monitoring}\label{sec:result_construction_testing:monitor}

\noindent\textbf{Limited tool support to continuously monitor the health status of software architectures} (\faCogs).
Architecture monitoring aims to quantify the health status of software systems in a continuous way, and evaluate whether the symptoms of architecture problems crept into a system~\cite{mirakhorli2015detecting}.
Despite the common awareness of the importance of architecture monitoring among our participants, few organizations employ specific automated tools for architecture monitoring in practice. As mentioned by the participants, without continuous architecture monitoring, teams would discover architecture problems at the late stages of software life cycle (P3a, P12a). When it comes to building software on top of legacy systems, it is difficult to make accurate project planning due to long-lived architecture problems in the legacy systems, as expressed by P3b (architect): ``\emph{the presence of weird  dependencies and calls in the legacy system makes it a headache to build the new system on top of it. This adds a bunch of risks when trying to make plans.}''

\noindent\textbf{Pinpointing architecture problems requires a system-wide perspective }(\faCalendar).
Pinpointing architecture problems with quantitative measures of architecture monitoring is difficult, as frequently mentioned in our interviews (P2a, P3a, P4c-d, P7a, P11a, P14a, P19b). As P2a (developer) illustrated, ``\emph{the root causes of performance degradation in architecture monitoring could arise from architecture problems, code issues, or both ... it is difficult to accurately pinpoint architecture problems that cause the performance degradation.}'' Pinpointing the root causes requires an overall perspective of the entire software system, yet individual developers tend to focus on the code of modules under development in the phase of software construction, thus lacking such perspective (P3a). 
The literature also discussed that the separation between interface and implementation in component-based software, along with the hiding of the implementation from the component client, could introduce problems that are difficult to pinpoint when their effects cross component boundaries~\cite{wuttke2010automatically}.

\noindent\textbf{Maintenance of monitoring tools requires ongoing effort} (\faGroup).
The maintenance of monitoring tools is another common challenge mentioned by participants.
In some organizations, automated testing frameworks are deployed to quantitatively monitor the performance of software systems, but cannot automatically adapt to the evolution of systems. 
The tools require significant effort and cost for their maintenance for expanding the scope of measured quality attributes and evolving thresholds of measurements (P3f, P4c, P5a, P12a).
For example, P3f (developer) shared, ``\emph{someone griped about the hassle of calibrating the quantitative measures as the system evolves.}''

\noindent\textbf{Recommendations.} 
Our observations suggest that the use of automated testing frameworks appears to be a practical way to continuously monitor architecture quality in terms of system performance (\faCalendar).
As our participants reported, some organizations use unit testing, stress testing, and performance testing frameworks, to continuously monitor the quality of software systems (P2a, P3a, P3c, P3e-f, P4b-d, P5a, P7a, P11a, P12a, P14a, P19b). The performance metrics generated by automated testing frameworks tend to serve as an indicator of architecture problems, e.g., performance degradation.

Participants raised several expectations for building tools to support continuous architecture monitoring (\faCogs): 
(1) integrating a vast variety of metrics for comprehensively evaluating software architectures, especially in terms of the quality attributes of software systems (P2a, P4a); (2) deriving the metrics from the software architecture experiences in practice, given architecture design is a human- and knowledge-intensive process (P3e-f); (3) backing up empirically based metrics with theoretical support (P1a); and (4) visualizing metrics to facilitate diagnose of potential architecture problems (P3b). In addition, it seems important to budget for the maintenance of architecture monitoring tools, or even a dedicated maintenance team (\faCalendar).

Involving architects is important when pinpointing architecture problems in continuous architecture monitoring (\faGroup). 
In the process of architecture monitoring, participants suggested involving architects to analyze potential problems indicated by monitoring tools, especially the potential problems in the core components of a software system (P2a, P3a, P3e). Conversely, developers are suggested to be actively involved (P3f), by providing explanations or rationale behind implementation decisions that violate the criteria for metrics as specified in monitoring tools. 

\subsection{Construction Quality}\label{sec:result_construction_testing:quality}
\textbf{Technical debts are introduced to software projects} (\faGroup,~\faCalendar).
Participants regard technical debts as a common cause of architecture erosion.
Technical debt is a metaphor reflecting technical compromises that yield short-term benefits but hurt the long-term success of software systems~\cite{tom2013exploration}. 
Participants reported a variety of technical debts in the course of software construction, including workarounds, shortcuts and sub-optimal operations (P2a, P3b, P3g, P4a-c, P5a, P6a-b, P7a, P11a, P12a, P13a, P14a, P15a). 
The workarounds, shortcuts and sub-optimal operations tend to change architecturally relevant elements, such as classes, components, and modules, introduce undesired dependencies among these elements, and further break architectural integrity, which is observed in previous work (e.g.,~\cite{brunet2012evolutionary}). 
Participants frequently reported that pragmatism, prioritization and ignorance may contribute to the introduction of technical debts. An example of pragmatism, creating a \emph{minimum viable product} in a short amount of time, is noted in prior research~\cite{klotins2018exploration,tom2013exploration}, which is also recounted by some participants (P3b, P3g, P5a, P6a). 
The implementation of critical functions is generally prioritized above architecture quality due to the fast pace and time pressure in software development (P2a, P4a-b). 
Ignorance refers to the inability of individual developers to construct high-quality software systems due to the lack of adequate knowledge on writing clean code, the applied technologies and business domains (P3d), which is also highlighted in a prior study~\cite{kruchten2019managing}. P3d (architect) illustrated,  ``\emph{without a deep understanding of a programming language or technology, developers tend to misuse the features of the language or technology when performing specific programming tasks ... unintentionally incur technical debt}.''

A few participants mentioned some examples of technical debts relating to design activities (P2a, P3b), including upfront detailed design of modules with an under-focus on quality attributes of software systems, such as maintainability and extensibility, and sub-optimal upfront solutions in software architectures.

\noindent\textbf{Lack of tool support for detecting architectural smells} (\faCogs).
Most participants notice the occurrence of architectural smells in their software systems, but lack tool support for detecting architectural smells (P2a, P3a-g, P4a-d, P6a-b, P11a, P12a, P13a, P15a).
Architectural smells indicate the structural problems in the components and their interactions with other components of software systems that are caused by architecture antipatterns, misuse or violation of architecture styles, and violation of design principles~\cite{mumtaz2021systematic}. 
Dependency related architectural smells are the most frequently mentioned architectural smells in our interviews (P3a-b, P3d, P3f-g, P4a, P4c-d, P11a, P19b), including cyclic, undesired and unstable dependencies~\cite{greifenberg2015architectural}. Some participants gave examples of cyclic dependency smells (P3b, P3d, P3f, P4b), as shared by P3b (architect) ``\emph{[cyclic dependency smells] happen when architecture components, such as modules or subsystems, depend on each other in some way, directly or indirectly}.''
The subsystems and microservices involved in a dependency cycle tend to be impractical or even impossible to separately release, deploy and maintain~\cite{azadi2019architectural}. 
The undesired dependency occurs when an architecture component depends on an excess number of other components (P3d), or the dependency introduces violations of design principles, e.g., top-down dependencies in layered architectures (P3g). In terms of unstable dependency, P3g (project manager, architect) mentioned that ``\emph{some modules depends on other modules that are less stable than itself ... [because of the dependencies] the more stable modules tend to change frequently with the less stable ones}''.

Modularity violation architectural smells are also frequently discussed in our interviews (P2a, P3a-b, P3d-g, P6a, P11a, P19b). 
Participants reported that software systems become increasingly complex as the development proceeds, with gradually losing cohesion in architectural components, thus deteriorating the modularity of the systems. 
For example, some participants observed \emph{god components} as an architectural smell~(P2a, P3b, P3d-f)~\cite{azadi2019architectural}, as expressed by P3b (architect): ``\emph{some components have taken on way too many responsibilities and completely disregarded the principle of separation of concerns}.''
The presence of modularity related architectural smells may imply the sub-optional decisions for decomposition and modularization in the design phase (P2a, P3a, P3d, P3f-g, P6a).

\noindent\textbf{Unawareness of correlation between code smells and architecture problems}(\faGroup).
Most organizations use a variety of code analysis tools for code smell detection, yet participants do not understand whether relationships exist between code smells and architecture problems. 
Specifically, organizations apply homegrown, open-source, or commercial tools for automated detection of code smells, including duplicated code (P3c, P3f, P11a, P12a, P17a), cyclomatic complexity (P3b), bad naming (P2a, P3b-d, P3f, P4b), and large class (P2a, P3b, P3e-f).
For example, P3c (developer lead, testing) shared, ``\emph{we've got tools to help us catch duplicate code and other code smells, but it's not clear how the results actually help us identify architecture problems}.''
Previous studies indicate that certain patterns of co-occurrence of code smells tend to be effective indicators of architectural erosion~\cite{li2011case,guimaraes2015architecture,macia2012automatically,bertran2011detecting}. 
Several code smell detection techniques look at the correlation between code smells and architectural smells for locating architecture problems (e.g.,~\cite{lenhard2017code}). Some code smell detection tools also employ code smell metrics that are relevant to the identification of architectural smells~\cite{lenhard2017code,macia2012relevance}.

\noindent\textbf{Recommendations.} 
Investing effort and time in addressing technical debts is important (\faGroup, \faCalendar). 
Some participants recommended employing commit-level refactoring to gradually reduce technical debts that contribute to architecture problems (P4a, P5a). A few participants suggested incremental benefit analysis for code changes relating to technical debts, which may encourage developers to prioritize development tasks that address technical debts.

It is important to automate the detection of architectural smells to facilitate the enforcement of architectural constraints in the construction phase (\faCogs, \faCalendar). Most participants emphasized that the automated detection of architectural smells, as a proactive prevention strategy of architecture erosion, can lower the cost of future maintenance for software systems (P2a, P3a-g, P4a-b, P6a). 
some participants recommended integrating architectural smell detection into the code review process, and involving architects to review code commits that incur architectural smells; developers are allowed to explain their code that incurs architectural smells (P3b).
To increase the usability of such tools, participants recommend highlighting affected code by architectural smells and providing actionable mitigation strategies (P3g). 

Detecting organizational-wide dependency related architectural smells requires the collaboration of multiple project teams (\faGroup, \faCogs).  
The detection of dependency related architectural smells requires the construction of dependency graphs between architectural components~\cite{azadi2019architectural}.
In some organizations, the architectural components are developed and maintained by multiple project teams, as some participants mentioned (P3a, P4a).
Thus, it is important to coordinate multiple project teams for the construction of dependency graphs.  \section{Software Maintenance}\label{sec:result_maintenance}
Once a software system is delivered and in operation, defects are uncovered, operating environments change, and new requirements surface, as the system evolves. Software maintenance aims to modify the software system and ensure that the system continues to satisfy requirements while preserving its integrity. 

\subsection{Architecture Erosion}\label{sec:refactoring_erosion}
Software architecture may exhibit an eroding tendency when changes are accumulated in a software system. As the system evolves, the accumulation of such problems can cause the implemented architecture to deviate away from the intended architecture. The phenomenon of divergence between the intended and implemented architectures is regarded as architecture erosion~\cite{li2022understanding}. 

\noindent\textbf{Lack of tool support to explicitly capture and aggregate symptoms of architecture erosion} (\faCogs).
Architecture erosion tends to affect the quality and evolution of software systems, manifesting various symptoms as mentioned in our interviews and the literature~\cite{li2022understanding}.
Nonetheless, automated tools are rarely used by practitioners in most organizations to explicitly capture and aggregate architecture erosion symptoms in terms of quality attributes as well as maintenance and evolution activities.

Most participants have perceived architecture erosion from the quality perspective, including performance degradation, frequent system failures, and high bug rate. In contrast, a few participants pointed out that software systems with eroded architectures might have a good runtime performance because ``\emph{[maintenance] teams only monitor limited metrics of runtime performance of key business requirements}'' (P4a; project manager, architect). 

Some participants have perceived architecture erosion from the perspective of maintenance and evolution activities, including difficulty in integrating new requirements into software systems, and locating and fixing bugs (P1a, P2a, P3a, P3b, P3f, P4a, P6a, P11a,  P13a, P17a). 
The difficulty in such activities manifests itself in a variety of ways, e.g., increased cost, complex implementation, and highly scattered code changes. For instance, P6a (developer) illustrated, ``\emph{[when a system suffers from architecture problems],  integrating requirements tends to cost more time and effort than expected based on experience}.''
Integrating a new requirement in an eroded architecture tends to be extremely complex, even for a simple business requirement (P1a, P3a-b, P3f). P4a (project manager, architect) gave an example of locating bugs in an eroded system with microservices architecture, ``\emph{a bug could arise from the code that spans across multiple microservices, making it hard to locate}.''
Moreover, bug fixing in an eroded architecture tends to involve highly scattered code changes, which affects an excessive number of files. P3f (developer) identified the risks of highly scattered code changes, ``\emph{[scattered code changes] are prone to bugs and failures in the future}.''

\noindent\textbf{Obsolete documentation and increasing complexity of software systems accelerate architecture erosion} (\faCalendar, \faGroup).
Obsolete documentation and increasing complexity of software are expressed as the top reasons that cause architecture erosion in our interviews (P1a, P3b, P3f, P4a-b), which have been also raised in the literature~\cite{jaafar2013relationship,de2012controlling,merkle2010stop,macia2012relevance}. Sometimes, practitioners implemented new requirements without updating the design documentation due to deadline pressure (P4b). 
Participants also attribute architecture erosion to the increasing complexity of architecture as a system evolves. 
Consequently, increasing complexity of software systems tends to reduce the understandability of architectures, and further deteriorates the architectures by sub-optimal implementations when changes occur~\cite{de2012controlling}, making the architecture ``\emph{cumbersome (heavy), complicated and fragmented}'' (P1a, P3c, P3e). 
For instance, P3d (architect) explained, ``\emph{it is impossible to integrate new requirements into a software system without degrading the quality of its architecture}.''

\noindent\textbf{Recommendations.} 
A few participants suggested building a dashboard to visualize architecture erosion symptoms, which integrates data from multiple sources, e.g., the number of bugs from issue tracking systems, runtime logs from the execution environment, and performance metrics from monitoring tools (P1a; \faCogs). 
\subsection{Architecture Refactoring}\label{sec:result_maintenance:refactor}

Architecture refactoring~\cite{ivers2022industry}, also known as large-scale refactoring, involves structural and broad changes at an architectural scale to maintain the structural quality of an evolving software system, and facilitate its success to facilitate the integration of new features. 

\noindent\textbf{No agreement on the value of architecture refactoring} (\faGroup).
Practitioners in organizations cannot reach an agreement on prioritizing architecture refactoring tasks over development tasks for delivering business products (P2a, P3a, P3e-f, P4a, P6a, P11a, P13a).
Despite the benefits of architectural refactoring, prior studies~\cite{ernst2015measure,kim2012field} also reported that developers perceive the structural changes at architectural scale as costly, complicated, and risky, and failing to implement such changes incurs significant consequences. 
The value of continuous maintenance activities relating to software architectures tends to be underestimated, because the return on investment is usually unclear (P6a). 
As P2a (developer) explained, ``\emph{given the potentially high anticipated cost of architecture refactoring, the senior management level would like to see clear quantifiable value [from architecture refactoring] for the organization}.''
Architecture refactoring is primarily business driven, including requirements for significant performance upgradation and delivery of new features that are not supported by existing architecture (P1a, P2a, P3g, P4a, P5a, P6a). 
In addition, technical reasons for architecture refactoring are also mentioned by some of the participants (P1a, P4b, P5a, P6a), including migration to new technologies and modernization. 

\noindent\textbf{Inadequate tool support for impact analysis of architectural changes} (\faCogs).
Participants perceive impact analysis as a challenging activity when it comes to architecture refactoring. 
Impact analysis for architecture refactoring requires a conceptual grasp of the overall code structure, as well as a deep understanding of design decisions of a software system, as discussed in the literature (e.g.,~\cite{ivers2022industry}).
P1a (architect) emphasized that ``\emph{impact analysis of architecture refactoring becomes even challenging for an aging system due to its increasing complexity and staff turnover, because no one in the team could be capable of performing accurate impact analysis }.''

Organizations rarely provide tool support for impact analysis of architecture refactoring due to limited traceability information between architecture, design and implementation artifacts. 
In addition, as a software system evolves over time, the traceability information becomes obsolete due to the separate evolution of its architecture and source code (P3b, P3e-g, P4a, P6a) (Section \ref{sec:result_construction_testing:conformance}; also observed in previous studies, e.g., ~\cite{javed2014systematic}).

\noindent\textbf{Inadequate tool support for module- and system-level refactoring} (\faCogs).
In practice, participants reported that their teams perform software refactoring of various granularities at different frequencies, local refactoring occasionally (p3a, P3e-f, P4a-b, P5a, P6a), module-level refactoring on a monthly basis (P1a, P3a, P3f, P4a, P5a), and system-level on a yearly basis (P1a, P3e, P11a). 
Most organizations integrate automated tools into their software engineering process to support local refactoring, but do not provide tool support for module- and system-level refactoring. 
Practitioners tend to manually perform module- and system-level refactoring, with limited support from automated testing frameworks. P3a compared automated testing frameworks to ``\emph{safety net}'' of a software system, which validates various aspects of artifacts under refactoring, e.g., functionality, performance, and user experience.
Prior studies also reported that software architects and developers primarily rely on manual efforts for large-scale refactoring activities, with minor support from disjoint tools~\cite{vakilian2012use,murphy-hill2011we,ivers2022industry}. 

\noindent\textbf{Recommendations.}
It is important to cultivate a software engineering culture in which the whole organization shares a common commitment to high-quality software architecture (\faGroup). Senior management and practitioners with different roles need to understand the implications of software architecture practice and their stakes in the system (P3a, P3e-f, P13a). 
Despite the importance of impact analysis in architecture refactoring there is little tool support due to limited traceability information (\faCogs, \faFileText). Impact analysis supports the identification of areas affected by possible changes (P3a, P3e, P4a, P5a), and the estimation of resources needed to accomplish the refactoring (P1a). 
In practice, some organizations encourage practitioners manually build traceability links by including snippets of design documentation or its link in commit description (P2a), Javadoc comments (P4b), or code review comments (P6a). The strategy enables the capturing of traceability links but requires significant maintenance efforts as architecture and source code evolves.  
A few organizations use a DevOps dashboard to build traceability links between user stories, design artifacts, and source code.
It seems beneficial to adopt a more formal process for building traceability across design and implementation artifacts (\faCalendar).  
\section{Discussion}\label{sec:discussion}

In the interviews, we observed that large-scale organizations tend to adopt better software architecture practices as compared to smaller organizations. While most organizations struggle to establish effective processes and tooling for achieving high-quality software architectures, smaller organizations struggle more, and have limited advice to draw from for improvement. Apart from the size of organizations, our findings indicate that architectural styles have a notable influence on software architecture practices, specifically in terms of documentation and the application of design principles. Interface specifications enable communication between subsystems in layered architectures, but tend to be limited at the microservice level in microservices architectures (Section \ref{sec:result_design_doc}). Layered architectures prioritize vertical decomposition for hardware independence, whereas microservice architectures emphasize horizontal decomposition based on business requirements (Section \ref{sec:result_design_principles}). In addition, participants highlighted the challenge of locating bugs in eroded systems with microservices architectures, as bugs can arise from code spanning across multiple microservices, posing difficulties in identification (Section \ref{sec:refactoring_erosion}). 

We observed that architects (17 participants) raised more challenges compared to non-architects (15 participants), with a median of 10 vs. 7 challenges. Similarly, participants with 10 or more years of experience (14 participants) raised more challenges than those with less than 10 years of experience (18 participants), with a median of 10 vs. 7 challenges. Regarding the product domains, participants from PaaS (1 participant), communications (2 participants), and intelligent device (4 participants) domains identified the highest number of challenges, with median values of 17, 12, and 12 challenges mentioned in the interviews, respectively.

Supplement F presents a summary of the challenges, along with the recommendations identified through our interviews.
Most organizations face common challenges in performing software architecture activities, in terms of four aspects: (1) collaboration and management support (\faGroup), (2) architecture knowledge management (\faFileText), (3) adoption, development and maintenance of tools (\faCogs), and (4) process for architecture design and evolution (\faCalendar).
Beyond the specific challenges discussed throughout this paper, we observe four broad themes that would benefit from more attention both in  practice and future research:

\faGroup~\textbf{Management:}
It is important for the strategic management level and practitioners in an organization to share a mindset that high-quality software architecture requires a continuous and significant investment of time and resources throughout the software life cycle, and benefits all stakeholders in software development and maintenance.  
The architects from different projects or business units can work in a cross-functional team, in which they collaborate in architecture evaluation and review, collect feedback from developers across projects to analyze architecture problems (e.g., architecture inconsistency and architectural smells), and capture emerging aspects of software architecture as systems evolve. 

\faFileText~\textbf{Documentation:}
Organizations use a variety of tools to capture and share architecture documentation, yet practitioners express the need for better modeling tools, versioning mechanisms, and lightweight support for traceability in practice (Section \ref{sec:result_design_doc}). Despite abundant research on architecture knowledge management~\cite{capilla2016ten}, little is used in practice to ensure the completeness, up-to-dateness, and traceability of architecture documentation. 
Thus, future research could put effort into designing a unified engineering tool for architecture knowledge management within an organization with support for sophisticated reasoning as well as evolution and runtime decision-making. 

\faCogs~\textbf{Tooling:}
Practitioners struggled with inadequate use or lack of automation and tool support for architecture conformance checking (Section \ref{sec:result_construction_testing:conformance}), continuous monitoring (Section \ref{sec:result_construction_testing:monitor}), architectural smell detection (Section \ref{sec:result_construction_testing:quality}), and architecture refactoring (Section \ref{sec:result_maintenance:refactor}). 
The development of such tools is a highly sophisticated and time-consuming process involving interdisciplinary expertise in both technical and business aspects of software systems.  
Some participants recommend building a dedicated team for tool development and maintenance, to alleviate additional burdens for development teams.
Traceability links between various software artifacts play a fundamental role in such tools, but require manual maintenance in practice which can be time consuming and error prone. Recent studies on automatic traceability recovery~\cite{aung2020literature} may provide support to the maintenance of traceability links.
Future research could investigate the applicability of such traceability recovery techniques in practice.

\faCalendar~\textbf{Process:}
In practice, architects tend to base the reasoning process of design decisions on their own experiences and expertise. Even though some organizations encourage the practice of including design decisions and rationale in architecture documentation, few provide guidance for representing design decisions and rationale to support the consumption of such architecture knowledge. We have seen a number of meta-models that aim at representing design decisions and rationale~\cite{li2013application,tofan2014past}. It can be beneficial to integrate the use of meta-models in the decision-making process for architecture design.

 \section{Threats to Validity}\label{sec:threat}

 Our study exhibits typical threats for qualitative research. Generalizations beyond the sampled participant distribution should be made with care.

Another threat could be the representativeness of our study demographics for the software industry in general, as the participants have been sampled through our personal networks. In several organizations, we only interviewed a single person, possibly giving us a one-sided perspective.
To mitigate this threat, we applied maximum variation sampling method to cover practitioners with varying levels of experience and diverse job roles, and from organizations of different sizes, product domains, and geographical regions. The purposive sampling and constant comparisons in data analysis lead to data saturation being achieved with a relatively small sample size of 32 interview participants.

The diverse range of organizations and experience of participants presented an opportunity to triangulate the overall dimensions of the broad topic of software architecture in practice. Despite the specialized experience within different organizations among various participants, the participants shared significant similarities in the challenges they encounter as evidenced in the follow-up validity check.  

\section{Conclusion and Future Work}\label{sec:conclusion}
This work followed a qualitative research strategy through interviews with 32 participants from 21 organizations to explore how practitioners perform software architecture related activities, as well as what challenges they face.
We observed significant variations among organizations in terms of their strategies, processes, and tools related to software architecture. Yet, we identified common challenges in performing software architecture activities, in terms of four aspects: (1) collaboration and management support, (2) architecture knowledge management, (3) adoption, development and maintenance of tools, and (4) process for architecture design and evolution. 
Future research could put efforts into quantitatively exploring how cultural, product, and company characteristics influence the software architecture practices and challenges, and further designing strategies to attain  high-quality software architectures throughout software development and maintenance.

\balance

\bibliographystyle{ACM-Reference-Format_abbrev}
\bibliography{main}

\pagebreak
\appendix
\section*{Supplementary Material}
\renewcommand{\thesubsection}{Supplement \Alph{subsection}}

\subsection{Codebook}\label{supplement:codebook}
\noindent\textbf{1. Software Requirements}

\noindent$\blacktriangleright$ \emph{Description and Relation to Software Architecture}: 
Software requirements process is concerned with the elicitation, analysis, specification, and validation of software requirements as well as the management of requirements. Architectural design is the point at which the requirements process overlaps with software design. Thus, software architecture practice would happen in the software requirements stage of a waterfall-like process or iteratively in other process models. 
Software requirements express the needs and constraints placed on a software system that contribute to the solution of some real-world problem. Some requirements, particularly certain nonfunctional ones, have a global scope in that their satisfaction cannot be allocated to a discrete component in a software system. Hence, a requirement with global scope may strongly affect the software architecture and the design of many components.

\noindent$\blacktriangleright$ \emph{Challenges (Code, Description and Example)}:

- Unpredictable requirements: As software systems evolve, the evolution and changes in software requirements, including new features and non-functional requirements.
\emph{Example}: 
``Architects cannot always foresee future requirements when they design architecture''.

\noindent\textbf{2. Software Design}

\noindent$\blacktriangleright$ \emph{Description and Relation to Software Architecture}: During software design, practitioners produce various models that form a kind of blueprint of the solution to be implemented. These models can be analyzed and evaluated to determine whether or not they will fulfill requirements. Practitioners also examine and evaluate alternative solutions and tradeoffs. Finally, the resulting models are used to plan subsequent development activities, such as using them as inputs and as the starting point of construction and testing. 
In a waterfall-like process or iteratively in other process models, software design consists of two activities that fit between software requirements and software construction: (1) Software architectural design, sometimes called high-level design, in which top-level structure and organization of a software system is produced,  and (2) software detailed design, in which each component is specified in sufficient detail to facilitate its construction.

An architectural style is a specialization of element and relation types, together with a set of constraints on how they can be used,  thus providing the high-level organization of software systems. A number of architectural styles have been identified:

- General structures (e.g., layered, pipes and filters, blackboard)

- Distributed systems (e.g., client-server, three-tiers, broker, microservices)

- Interactive systems (e.g., Model-View-Controller, Presentation-Abstraction-Control)

- Adaptable systems (e.g., microkernel, reflection)

- Others (e.g., batch, interpreters, process control, rule-based)

\emph{\textbf{2.1 Design Documentation}}

\noindent$\blacktriangleright$ \emph{Description and Relation to Software Architecture}: Different high-level views of a software design can be described and documented. A view represents a partial aspect of a software architecture that shows specific properties of a software system. For instance, the logical view represents functional requirements, the process view represents concurrency issues, the physical view represents distribution issues, and the development view represents how the design is broken down into implementation elements with explicit representation of the dependencies among the elements.

\noindent$\blacktriangleright$ \emph{Challenges (Code, Description and Example)}:

- Completeness: Use of models and tools is  inadequate to ensure the completeness of architecture documentation. \emph{Example:} ``I cannot find the relevant information in architecture documentation''.

- Up-to-dateness: Architecture documentation becomes obsolete as software evolves. \emph{Example:} ``Documentation-code inconsistency sometimes confuses me when I implement new features [for the system]''.

- Conflict with process: Conflict between documentation and agile process. \emph{Example}: ``Traditional modeling method like UML becomes a significant roadblock to fast-paced design in agile development''.

- Inadequate tool support: Inadequate tool support for sharing, version control, and tracing of scattered design documentation. \emph{Example}: ``Our company uses text processing software to collect  design documentation, and version control systems to keep track
of changes in documents''.

\emph{\textbf{2.2 Design Principles}}

\noindent$\blacktriangleright$ \emph{Description and Relation to Software Architecture}: Software design principles provide the basis for many different software design approaches and concepts, including abstraction, coupling and cohesion, decomposition and modularization, encapsulation/ information hiding, separation of interface and implementation, sufficiency, completeness, and primitiveness, and separation of concerns.

\noindent$\blacktriangleright$ \emph{Challenges (Code, Description and Example)}: 

- Software decomposition: Unclear boundaries between architectural elements in software systems.  
\emph{Example}: ``One cannot elegantly decompose a software system by simply collecting and listing a bunch of business scenarios''.

- Interdisciplinary knowledge: Interdisciplinary knowledge is required to lower coupling and improve cohesion of software.
\emph{Example}: ``Business requirements change over time in different frequencies ... a component [in a software system] tend to be highly coupled with others [as the system evolves] if it is responsible for both frequently and rarely changed requirements''.

\emph{\textbf{2.3 Design Quality Analysis and Evaluation}}

\noindent$\blacktriangleright$ \emph{Description and Relation to Software Architecture}: Various attributes contribute to the quality of a software design, including quality attributes discernible at runtime (e.g., performance,
security, availability, functionality, usability) and those not discernible at runtime (e.g., modifiability, portability, reusability, testability).
Various tools and techniques can help in analyzing and evaluating software design quality, including software design reviews to determine the quality of design artifacts, static analysis to detect design errors, as well as simulation and prototyping to evaluate a design.
Measures can be used to assess or to quantitatively estimate various aspects of a software design, e.g., size, structure, or quality.

\noindent$\blacktriangleright$ \emph{Challenges (Code, Description and Example)}:

- Lack of standard process and tool support: Architecture review requires a standard process, active involvement of external experts, and tool support. 
\emph{Example}: ``Informal architecture reviews would take place once every one to two years during software maintenance, planning for the evolution of software architectures''. 

- Experience based: Evaluation and review of architectures heavily rely on practitioners' experiences and expertise. 
\emph{Example}: ``Experts rely on their experience in architecture review and evaluation process when inspecting the requirements and potential solutions in architecture documentation''.

- Lack of apply-to-all measures: Effective measurements differ across software systems. \emph{Example:} ``[Our company] implemented quantitative measures to evaluate architecture quality, but they are superficial and lack of theoretical basis''.

\noindent\textbf{3. Software Construction and Testing}

\noindent$\blacktriangleright$ \emph{Description and Relation to Software Architecture}: Software construction refers to the creation of working software through a combination of coding, verification, unit testing, integration testing, and debugging; Software testing consists of dynamic verification of expected behaviors in software systems. The construction and testing phases use design output. Boundaries between design, construction, and testing would vary depending on the processes used in software projects.

\emph{\textbf{3.1 Architecture Conformance Checking}}

\noindent$\blacktriangleright$ \emph{Description and Relation to Software Architecture}: Architecture conformance checking examines whether the implemented architecture is consistent with the intended architecture, ensuring developers have followed the architectural edicts set and not eroding the architecture by breaking down abstractions, bridging layers, and compromising information hiding.

\noindent$\blacktriangleright$ \emph{Challenges (Code, Description and Example)}: 

- Lack of tool support: Practitioners rely heavily on periodical manual inspection for architecture conformance.  
\emph{Example}: ``Detailed description [of architecture-level code changes] would save the time of architects to evaluate whether the code changes conform with the intended architecture''.

- Obsolete documentation: Obsolete documentation hinders the  automation of architecture conformance checking.
\emph{Example}: ``Developers tend to forget to update architecture documentation when evolving software architectures because of deadline pressures''.

- Lack of traceability: Lack of traceability hinders the automation of architecture conformance checking. 
\emph{Example}: ``No standard process or tool support exists to build trace links between design decisions and their implementation''.

\emph{\textbf{3.2 Architecture Monitoring}}

\noindent$\blacktriangleright$ \emph{Description and Relation to Software Architecture}: Architecture monitoring aims at employing means to monitor the health status of software systems, and evaluate whether architecture erosion symptoms crept into implemented software architecture.

\noindent$\blacktriangleright$ \emph{Challenges (Code, Description and Example)}: 

- Limited tool support: Limited tool support to continuously monitor the health status of software architectures.
\emph{Example}: ``We usually identified architecture problems at the late stages of software life cycle, with no support from continuous architecture monitoring''.

- System-wide perspective: Pinpointing architecture problems requires a system-wide perspective.
\emph{Example}: ``The root causes of performance degradation in architecture monitoring could arise from architecture problems, code issues, or both ... it is difficult to accurately pinpoint architecture problems that cause the performance degradation''.

- Maintenance effort for tooling: Maintenance of monitoring tools requires ongoing effort.
\emph{Example}: ``We use automated testing frameworks to quantitatively monitor the performance of software systems, but [the frameworks] cannot automatically adapt to the evolution of systems''.

\emph{\textbf{3.3 Construction Quality}}

\noindent$\blacktriangleright$ \emph{Description and Relation to Software Architecture}: Faults can be introduced during construction and result in serious quality problems. Numerous techniques exist to ensure the quality of code as it is constructed, including static analysis (e.g., code smell detection), unit testing, and integration testing.
Technical debt reflects technical compromises that yield short-term benefits but hurt the long-term success of software systems. It might give rise to suboptimal design decisions, and cause architecture erosion with a high probability.
Architectural smells indicate the structural problems in the components and their interactions with other components of software systems that are caused by architecture antipatterns, misuse or violation of architectural styles, and violation of design principles.
Certain patterns of co-occurring code smells tend to be strong indicators of architecture erosion.

\noindent$\blacktriangleright$ \emph{Challenges (Code, Description and Example)}: 

- Technical debts: Technical debts are introduced to software projects, which is regarded as a common cause of architecture erosion.
\emph{Example}: ``Without a deep understanding of a programming language or technology, developers tend to misuse the features of the language or technology when performing specific programming tasks ... unintentionally incur technical debt''.

- Lack of tool support: Lack of tool support for detecting architectural smells. 
\emph{Example}: ``[Unstable dependency] ... some modules depend on other modules that are less stable than itself ... [because of the dependencies] the more stable modules tend to change frequently with the less stable ones''.

- Unawareness of architecture problems: Unawareness of correlation between code smells and architecture problems.
\emph{Example}: ``We use tools for code smell detection, but it is unclear how code smells relate to architecture problems''.

\noindent\textbf{4. Software Maintenance}

\noindent$\blacktriangleright$ \emph{Description and Relation to Software Architecture}: Software maintenance sustains the software system from development to operations throughout its life cycle. Software development efforts result in the delivery of a software system that satisfies user requirements. Once a software system is delivered and in operation, defects are uncovered, operating environments change, and new user requirements surface. Software maintenance aims to modify existing system and ensure that the system continues to satisfy user requirements while preserving its integrity. Four categories of maintenance have been defined: corrective, adaptive, perfective, and preventive maintenance.

\emph{\textbf{4.1 Architecture Erosion}} 

\noindent$\blacktriangleright$ \emph{Description and Relation to Software Architecture}: Software architecture may exhibit an eroding tendency when changes are accumulated in a software system. As the system evolves, the accumulation of such problems can cause the implemented architecture to deviate away from the intended architecture. The phenomenon of divergence between the intended and implemented architectures is regarded as architecture erosion. An eroded architecture can aggravate the brittleness of the system and decrease architecture sustainability.

\noindent$\blacktriangleright$ \emph{Challenges (Code, Description and Example)}: 

- Lack of tool support: Lack of tool support to explicitly capture and aggregate symptoms of architecture erosion.
\emph{Example}: ``[Maintenance] teams only monitor limited metrics of runtime performance of key business requirements''.

- Obsolete documentation: Obsolete documentation accelerates the erosion of software  architectures. 
\emph{Example}: 
``The outdated design documentation could hinder knowledge transfer, and cause a poor understanding of project contexts among the members of development teams''.

- Increasing complexity of software systems: Increasing complexity of software systems accelerates the erosion of software  architectures. 
\emph{Example}: ``Increasing complexity of software systems reduces the understandability of architectures, and deteriorates the architectures by sub-optimal implementations when changes occur, making the architecture cumbersome, complicated and fragmented''.

\emph{\textbf{4.2 Architecture Refactoring}}

\noindent$\blacktriangleright$ \emph{Description and Relation to Software Architecture}: Refactoring is a reengineering technique that aims at reorganizing a software system without changing its behavior to improve its structure and maintainability. Periodic architecture refactoring is required to maintain the structural quality of a complex and evolving software system, and for its success to facilitate the integration of new features. Other related terms, such as system-wide refactoring is also used in the literature.  Architecture refactoring usually takes a long time with considerable planning and team coordination involved.

\noindent$\blacktriangleright$ \emph{Challenges (Code, Description and Example)}: 

- Undervalued: No agreement on the value of architecture refactoring. 
\emph{Example}: ``Given the potentially high anticipated cost of architecture refactoring, the senior management level would like to see clear quantifiable value [from architecture refactoring] for the organization''.

- Inadequate tool support for impact analysis: Inadequate tool support for impact analysis of architectural changes. 
\emph{Example}: ``Impact analysis of architecture refactoring becomes even challenging for an aging system due to its increasing complexity and staff turnover, because no one in the team could be capable of performing accurate impact analysis''.

- Inadequate tool support for architecture refactoring: Inadequate tool support for module- and system-level refactoring. 
\emph{Example}: ``We use automated testing frameworks as a safety net to validate various aspects of artifacts under refactoring like functionality, performance, and user experience''.

\subsection{Interview Guide}\label{supplement:guide}

\noindent\textbf{{\large Introduction}}

\begin{itemize}
\item Thanks for taking the time to meet with me today. We are conducting a study to explore software architecture practice in industry. In support of this study, we are conducting a series of targeted interviews to gather information. 
During the interview, I would like to ask you a few questions about your background, then we will discuss your experience with respect to software architecture. 

\item In any data collected, or in reports or papers that are published, you will not be identified by name. Please be careful not to discuss any sensitive information about your company you work for. If you do mention any, we will do our best to remove it from our transcripts, but better if you don't mention such sensitive information at all.

\item Before we begin, I would like to notify you that I intend to record the interview for transcription purposes. Only the principal investigator will be able to access the recordings. After the interview is transcribed and identifying information is removed from the transcript, the audio recording will be destroyed.
\end{itemize}

\noindent\textbf{{\large Interview Questions}}

\noindent\emph{[Demographics]}

\begin{itemize}

\item[] \textbf{Years of Experience} How many years of experience do you have in total with software development and maintenance? And how many years with software architecture?

\item[] \textbf{Tenure} How many years you have been with your current company/organization? What is the size and geography of the company/organization?

\item[] \textbf{Job Responsibility} Next, I'm going to name a series of job responsibilities, and I'd like to tell me where you have ``none'', ``some'' or ``extensive'' experience you have for each job responsibility: [architecture design, detailed design, implementation, management, testing]

\item[] \textbf{Programming Languages} What programming languages do you use in your work? What is the most frequently used?

\item[] \textbf{Major Project} Can you tell me about the major project you worked on at your company? What are your roles in this project? How long have you been involved in this project?

\end{itemize}

\noindent\emph{[Topic-Specific Questions]}

Please answer the following questions regarding the major project you have been involved in.

\begin{itemize}

\item[] \textbf{Architecture} Can you describe the architecture of the system? What kind of architectural style does the system use? In your opinion, what do you think high-quality architecture is? 

\textbf{Design Process} Can you describe a bit about how you and your team design the system? 
\begin{itemize}
    \item[] \textbf{Design Principles} What design principles do you consider when designing the system? 
    \item[] \textbf{Design Documentation} Do you develop design documentation? What is included in design documentation? How do you write and maintain design documentation? Do you maintain versioning?
    \item[] \textbf{Design Decisions} How do you and your team make design decisions? Do you record design decisions? Were the design decisions traceable to the code?
    \item[] \textbf{Architecture Review and Evaluation} How do you review and evaluate architecture? Who are involved?
\end{itemize}

\item[] \textbf{Construction and Testing Process} Have you and your team encountered any architecture issues when implementing and testing the system?

\begin{itemize}
    \item[] \textbf{Identification} How did you identify such architecture issues? Any tool support?
    \item[] \textbf{Elimination} How did you eliminate such architecture issues? Any tool support?
    \item[] \textbf{Root Cause} In your opinion, why did such architecture issues arise? 
\end{itemize}
    
\item[] \textbf{Maintenance Process} Have you and your team encountered any architecture issues when maintaining the system?

\begin{itemize}
    \item[] \textbf{Identification} How did you identify such architecture issues? Any tool support?
    \item[] \textbf{Elimination} How did you eliminate such architecture issues? Any tool support?
    \item[] \textbf{Root Cause} In your opinion, why did such architecture issues arise? 
\end{itemize}

\end{itemize}
\noindent\emph{[Last Thoughts]}

\textbf{Challenges and Solutions} Can you think about any other challenges you faced with respect to software architecture during project development and maintenance? What do you think the potential solutions would be to these challenges?
\subsection{Interview Participants}\label{supplement:interviewees}
Table~\ref{tab:participant_details} presents the details of interviewed participants, companies and products.

\begin{table*}[t]
    \centering
    \caption{Details of interviewed participants, companies and products.}
      \begin{tabular}{lllll}
      \toprule
      \textbf{Company ID} & \textbf{Company Type} & \textbf{Participant ID} & \textbf{Participant Role} & \textbf{Product Domain} \\
      \midrule
      C1    & Mid-size Tech & 1a    & Architect & SaaS \\
      \midrule
      C2    & Big Tech & 2a    & Developer & Deep Learning \\
      \midrule
      C3    & Big Tech & 3a    & Architect & Intelligent Device \\
            &       & 3b    & Architect & Intelligent Device \\
            &       & 3c    & Developer Lead, Testing & Productivity Tool \\
            &       & 3d    & Architect & IaaS \\
            &       & 3e    & Architect, Developer Lead & Communications \\
            &       & 3f    & Developer & Communications \\
            &       & 3g    & Project Manager, Architect & Intelligent Device \\
      \midrule
      C4    & Big Tech & 4a    & Project Manager, Architect & E-commerce \\
            &       & 4b    & Architect & E-commerce \\
            &       & 4c    & Project Manager, Architect & E-commerce \\
            &       & 4d    & Architect, Developer & E-commerce \\
      \midrule
      C5    & Non IT & 5a    & Architect & Finance \\
      \midrule
      C6    & Big Tech & 6a    & Developer & E-commerce \\
            &       & 6b    & Developer & E-commerce \\
      \midrule
      C7    & Mid-size Tech & 7a    & Architect, Developer & Firmware \\
      \midrule
      C8    & Non IT & 8a    & Project Manager & Finance \\
      \midrule
      C9    & Non IT & 9a    & Developer & Health \\
      \midrule
      C10   & Startup & 10a   & Project Manager & SaaS \\
      \midrule
      C11   & Big Tech & 11a   & Project Manager & PaaS \\
      \midrule
      C12   & Non IT & 12a   & Project Manager & Finance \\
      \midrule
      C13   & Mid-size Tech & 13a   & Project Manager, Architect & Health \\
      \midrule
      C14   & Startup & 14a   & Project Manager, Architect & Supply Chain \\
      \midrule
      C15   & Big Tech & 15a   & Developer & Productivity Tool \\
      \midrule
      C16   & Non IT & 16a   & Architect & Finance \\
      \midrule
      C17   & Mid-size Tech & 17a   & Architect & Intelligent Device \\
      \midrule
      C18   & Startup & 18a   & Project Manager, Developer, Testing & Supply Chain \\
      \midrule
      C19   & Big Tech & 19a   & Developer & E-commerce \\
            &       & 19b   & Developer & Social Media \\
      \midrule
      C20   & Non IT & 20a   & Architect & Automotive and Clean Energy \\
      \midrule
      C21   & Mid-size Tech & 21a   & Project Manager & Health \\
      \bottomrule
      \end{tabular}\label{tab:participant_details}\end{table*}
  
  \subsection{Architectural Styles}\label{sec:style} 
  
\begin{figure}[t]
    \center
    \includegraphics[scale=0.5]{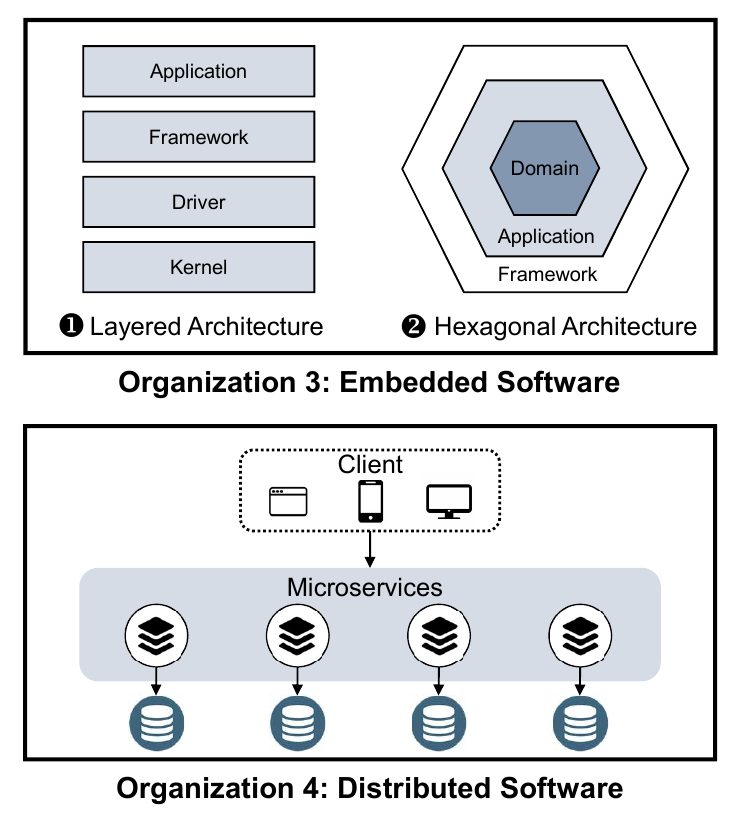}
    \vspace{-0.4cm}
    \caption{Architectural styles of two organizations.}
    \label{fig:styles} 
    \vspace{-0.3cm}
    \end{figure}

Throughout our interviews, we observed that the architectural styles for software systems to which our participants contribute differ widely across various application domains.
To illustrate the differences, we provide simplified descriptions of architectural styles applied in two application domains in Figure~\ref{fig:styles}. We show the building blocks and structures of architectural styles, as well as the responsibility of each building block.

Organization 3 (Figure~\ref{fig:styles}, top) develops embedded software on various devices for their clients, where participants reported two architectural styles applied to their products. The layered architecture is comprised of four layers: (1) the kernel layer, which implements application-specific customization of kernel functionality, (2) the driver layer, which initializes and manages access to the hardware devices, (3) the framework layer, which creates APIs over native libraries to simplify access to low-level components, and (4) the application layer, which delivers application-specific services to the users.

The hexagonal architecture is an alternative to the layered architectural style, which organizes the logical view in a way that places the business logic at the center.As shown in the hexagonal architecture in Figure~\ref{fig:styles}, the business logic is specified in the domain layer at the core of the architecture.
The application layer lies around the domain layer to isolate it from external factors and accomplish use scenarios.
The framework layer sits outside of the application layer, which implements services defined by the application layer.

Organization 4 (Figure~\ref{fig:styles}, bottom) develops distributed software systems to support their business objectives and activities, where participants reported microservices architectural style is applied to the server side of systems. 
In microservices architectures, the server side is decomposed into a set of microservices that communicate with each other through lightweight mechanisms, e.g., RESTful API or stream-based communications.Each microservice is implemented and operated as an independent unit, offering access to its internal logic and data through a well-defined network interface.For the client side, some systems can support a variety of clients including browsers, mobile applications, and desktop applications; others are middleware, which has no specific client side and only provides services to other systems or business applications in the organization.  
Despite the common use of microservices architecture on the server side, teams in Organization 4 develop software systems with different technology stacks.

\newpage
\subsection{Papers for Triangulation}\label{supplement:papers}
\begin{itemize}
            \item [1] Sandun Dasanayake, Sanja Aaramaa, Jouni Markkula, and Markku Oivo. Impact
            of requirements volatility on software architecture: How do software teams keep
            up with ever-changing requirements? Journal of software: evolution and process,
            31(6):e2160, 2019.

            \item [2] Marcin Szlenk, Andrzej Zalewski, and Szymon Kijas. Modelling architectural de-
            cisions under changing requirements. In 2012 Joint Working IEEE/IFIP Conference
            on Software Architecture and European Conference on Software Architecture, pages
            211–214. IEEE, 2012.

            \item [3] Lakshitha De Silva and Dharini Balasubramaniam. Controlling software ar-
            chitecture erosion: A survey. Journal of Systems and Software, 85(1):132–151,
            2012.

            \item [4] Tom Mens, Serge Demeyer, Olivier Barais, Anne Françoise Le Meur, Laurence
            Duchien, and Julia Lawall. Software architecture evolution. Software Evolution,
            pages 233–262, 2008.

            \item [5] Antony Tang, Peng Liang, and Hans Van Vliet. Software architecture documen-
            tation: The road ahead. In 2011 Ninth Working IEEE/IFIP Conference on Software
            Architecture, pages 252–255. IEEE, 2011.
            
            \item [6] Davide Arcelli, Vittorio Cortellessa, Antonio Filieri, and Alberto Leva. Control
            theory for model-based performance-driven software adaptation. In Proceed-
            ings of the 11th International ACM SIGSOFT Conference on Quality of Software
            Architectures, pages 11–20, 2015.
            
            \item [7] Benoit Baudry, Martin Monperrus, Cendrine Mony, Franck Chauvel, Franck
            Fleurey, and Siobhán Clarke. Diversify: ecology-inspired software evolution
            for diversity emergence. In 2014 Software Evolution Week-IEEE Conference on
            Software Maintenance, Reengineering, and Reverse Engineering (CSMR-WCRE),
            pages 395–398. IEEE, 2014.
            
            \item [8] Gerhard Fischer and J Otswald. Knowledge management: problems, promises,
            realities, and challenges. IEEE Intelligent systems, 16(1):60–72, 2001.
            
            \item [9] Muhammad Ali Babar and Ian Gorton. A tool for managing software architecture
            knowledge. In Second Workshop on Sharing and Reusing Architectural Knowledge-
            Architecture, Rationale, and Design Intent (SHARK/ADI’07: ICSE Workshops 2007),
            pages 11–11. IEEE, 2007.
            
            \item [10] G Maarten Bonnema. Communication in multidisciplinary systems architecting.
            Procedia CIRP, 21:27–33, 2014.
            
            \item [11] Bran Selic. Agile documentation, anyone? IEEE software, 26(6):11–12, 2009.
            
            \item [12] Junji Zhi, Vahid Garousi-Yusifoğlu, Bo Sun, Golara Garousi, Shawn Shahnewaz,
            and Guenther Ruhe. Cost, benefits and quality of software development docu-
            mentation: A systematic mapping. Journal of Systems and Software, 99:175–198,
            2015.
            
            \item [13] Emad Aghajani, Csaba Nagy, Olga Lucero Vega-Márquez, Mario Linares-Vásquez,
            Laura Moreno, Gabriele Bavota, and Michele Lanza. Software documentation
            issues unveiled. In 2019 IEEE/ACM 41st International Conference on Software
            Engineering (ICSE), pages 1199–1210. IEEE, 2019.
            
            \item [14] Wei Ding, Peng Liang, Antony Tang, and Hans Van Vliet. Knowledge-based ap-
            proaches in software documentation: A systematic literature review. Information
            and Software Technology, 56(6):545–567, 2014.
            
            \item [15] Philippe B Kruchten. The 4+1 view model of architecture. IEEE software, 12(6):42–
            50, 1995.
            
            \item [16] Andrew Forward and Timothy C Lethbridge. The relevance of software doc-
            umentation, tools and technologies: a survey. In Proceedings of the 2002 ACM
            symposium on Document engineering, pages 26–33, 2002.
            
            \item [17] Paul Clements, David Garlan, Reed Little, Robert Nord, and Judith Stafford.
            Documenting software architectures: views and beyond. In 25th International
            Conference on Software Engineering, 2003. Proceedings., pages 740–741. IEEE, 2003.
            
            \item [18] Marcel Rebouças, Renato O Santos, Gustavo Pinto, and Fernando Castor. How
            does contributors’ involvement influence the build status of an open-source
            software project? In 2017 IEEE/ACM 14th International Conference on Mining
            Software Repositories (MSR), pages 475–478. IEEE, 2017.
            
            \item [19] Rick Kazman, Dennis Goldenson, Ira Monarch, William Nichols, and Giuseppe
            Valetto. Evaluating the effects of architectural documentation: A case study of
            a large scale open source project. IEEE Transactions on Software Engineering,
            42(3):220–260, 2015.
            
            \item [20] Chung-Horng Lung, Marzia Zaman, and Amit Nandi. Applications of clustering
            techniques to software partitioning, recovery and restructuring. Journal of
            Systems and Software, 73(2):227–244, 2004.
            
            \item [21] Brian S Mitchell and Spiros Mancoridis. On the automatic modularization of
            software systems using the bunch tool. IEEE Transactions on Software Engineering,
            32(3):193–208, 2006.
            
            \item [22] Gabriela Şerban and István-Gergely Czibula. Object-oriented software systems
            restructuring through clustering. In Artificial Intelligence and Soft Computing–
            ICAISC 2008: 9th International Conference Zakopane, Poland, June 22-26, 2008
            Proceedings 9, pages 693–704. Springer, 2008.
            
            \item [23] Jian Feng Cui and Heung Seok Chae. Applying agglomerative hierarchical
            clustering algorithms to component identification for legacy systems. Information
            and Software technology, 53(6):601–614, 2011.
            
            \item [24] Qusay I Sarhan, Bestoun S Ahmed, Miroslav Bures, and Kamal Z Zamli. Software
            module clustering: An in-depth literature analysis. IEEE Transactions on Software
            Engineering, 48(6):1905–1928, 2020.
            
            \item [25] Shanshan Li, He Zhang, Zijia Jia, Zheng Li, Cheng Zhang, Jiaqi Li, Qiuya Gao,
            Jidong Ge, and Zhihao Shan. A dataflow-driven approach to identifying microser-
            vices from monolithic applications. Journal of Systems and Software, 157:110380,
            2019.
            
            \item [26] Rui Chen, Shanshan Li, and Zheng Li. From monolith to microservices: A
            dataflow-driven approach. In 2017 24th Asia-Pacific Software Engineering Confer-
            ence (APSEC), pages 466–475. IEEE, 2017.
            
            \item [27] Sara Hassan and Rami Bahsoon. Microservices and their design trade-offs: A self-
            adaptive roadmap. In 2016 IEEE International Conference on Services Computing
            (SCC), pages 813–818. IEEE, 2016.
            
            \item [28] Muhammad Ali Babar and Ian Gorton. Software architecture review: The state
            of practice. Computer, 42(7):26–32, 2009.
            
            \item [29] Ran Mo, Yuanfang Cai, Rick Kazman, Lu Xiao, and Qiong Feng. Decoupling level:
            a new metric for architectural maintenance complexity. In Proceedings of the 38th
            International Conference on Software Engineering, pages 499–510, 2016.
            
            \item [30] Santonu Sarkar, Avinash C Kak, and Girish Maskeri Rama. Metrics for measur-
            ing the quality of modularization of large-scale object-oriented software. IEEE
            Transactions on Software Engineering, 34(5):700–720, 2008.
            
            \item [31] Mikhail Perepletchikov, Caspar Ryan, and Keith Frampton. Cohesion metrics for
            predicting maintainability of service-oriented software. In Seventh International
            Conference on Quality Software (QSIC 2007), pages 328–335. IEEE, 2007.
            
            \item [32] Jianjun Zhao. On assessing the complexity of software architectures. In Proceed-
            ings of the third International Workshop on Software Architecture, pages 163–166, 1998.
            
            \item [33] Eric Bouwers, José Pedro Correia, Arie van Deursen, and Joost Visser. Quantifying
            the analyzability of software architectures. In 2011 Ninth Working IEEE/IFIP
            Conference on Software Architecture, pages 83–92. IEEE, 2011.
            
            \item [34] Dewayne E Perry and Alexander L Wolf. Foundations for the study of software
            architecture. ACM SIGSOFT Software engineering notes, 17(4):40–52, 1992.
            
            \item [35] Andrea Caracciolo, Mircea Filip Lungu, and Oscar Nierstrasz. A unified approach
            to architecture conformance checking. In 2015 12th Working IEEE/IFIP Conference
            on Software Architecture, pages 41–50. IEEE, 2015.
            
            \item [36] Andrea Caracciolo, Mircea Filip Lungu, and Oscar Nierstrasz. How do software
            architects specify and validate quality requirements? In European Conference on
            Software Architecture, pages 374–389. Springer, 2014.
            
            \item [37] Arthur Strasser, Benjamin Cool, Christoph Gernert, Christoph Knieke, Marco
            Körner, Dirk Niebuhr, Henrik Peters, Andreas Rausch, Oliver Brox, Stefanie
            Jauns-Seyfried, et al. Mastering erosion of software architecture in automotive
            software product lines. In International Conference on Current Trends in Theory
            and Practice of Informatics, pages 491–502. Springer, 2014.
            
            \item [38] Mehdi Mirakhorli and Jane Cleland-Huang. Detecting, tracing, and monitoring
            architectural tactics in code. IEEE Transactions on Software Engineering, 42(3):205–
            220, 2015.
            
            \item [39] Jochen Wuttke. Automatically generated runtime checks for design-level constraints.
            PhD thesis, Università della Svizzera italiana, 2010.
            
            \item [40] Edith Tom, Aybüke Aurum, and Richard Vidgen. An exploration of technical
            debt. Journal of Systems and Software, 86(6):1498–1516, 2013.
            
            \item [41] João Brunet, Roberto Almeida Bittencourt, Dalton Serey, and Jorge Figueiredo.
            On the evolutionary nature of architectural violations. In 2012 19th Working
            conference on reverse engineering, pages 257–266. IEEE, 2012.
            
            \item [42] Eriks Klotins, Michael Unterkalmsteiner, Panagiota Chatzipetrou, Tony Gorschek,
            Rafael Prikladnicki, Nirnaya Tripathi, and Leandro Bento Pompermaier. Explo-
            ration of technical debt in start-ups. In Proceedings of the 40th International
            Conference on Software Engineering: Software Engineering in Practice, pages 75–84,
            2018.
            
            \item [43] Philippe Kruchten, Robert Nord, and Ipek Ozkaya. Managing technical debt:
            Reducing friction in software development. Addison-Wesley Professional, 2019.
            
            \item [44] Haris Mumtaz, Paramvir Singh, and Kelly Blincoe. A systematic mapping study
            on architectural smells detection. Journal of Systems and Software, 173:110885,
            2021.
            
            \item [45] Fangchao Tian, Peng Liang, and Muhammad Ali Babar. How developers dis-
            cuss architecture smells? an exploratory study on stack overflow. In 2019 IEEE
            international conference on software architecture (ICSA), pages 91–100. IEEE, 2019.
            
            \item [46] Sunny Wong, Yuanfang Cai, Miryung Kim, and Michael Dalton. Detecting
            software modularity violations. In Proceedings of the 33rd International Conference
            on Software Engineering, pages 411–420, 2011.
            
            \item [47] Timo Greifenberg, Klaus Müller, and Bernhard Rumpe. Architectural consistency
            checking in plugin-based software systems. In Proceedings of the 2015 European
            Conference on Software Architecture Workshops, pages 1–7, 2015.
            
            \item [48] Umberto Azadi, Francesca Arcelli Fontana, and Davide Taibi. Architectural
            smells detected by tools: a catalogue proposal. In 2019 IEEE/ACM International
            Conference on Technical Debt (TechDebt), pages 88–97. IEEE, 2019.
            
            \item [49] Zude Li and Jun Long. A case study of measuring degeneration of software archi-
            tectures from a defect perspective. In 2011 18th Asia-Pacific Software Engineering
            Conference, pages 242–249. IEEE, 2011.
            
            \item [50] Everton Guimarães, Alessandro Garcia, and Yuanfang Cai. Architecture-sensitive
            heuristics for prioritizing critical code anomalies. In Proceedings of the 14th
            International Conference on Modularity, pages 68–80, 2015.
            
            \item [51] Isela Macia, Joshua Garcia, Daniel Popescu, Alessandro Garcia, Nenad Medvi-
            dovic, and Arndt von Staa. Are automatically-detected code anomalies relevant
            to architectural modularity? an exploratory analysis of evolving systems. In
            Proceedings of the 11th annual international conference on Aspect-oriented Software
            Development, pages 167–178, 2012.
            
            \item [52] Isela Macia Bertran. Detecting architecturally-relevant code smells in evolving
            software systems. In Proceedings of the 33rd International Conference on Software
            Engineering, pages 1090–1093, 2011.
            
            \item [53] Jörg Lenhard, Mohammad Mahdi Hassan, Martin Blom, and Sebastian Herold.
            Are code smell detection tools suitable for detecting architecture degradation? In
            Proceedings of the 11th European Conference on Software Architecture: Companion
            Proceedings, pages 138–144, 2017.
            
            \item [54] Isela Macia, Roberta Arcoverde, Alessandro Garcia, Christina Chavez, and Arndt
            Von Staa. On the relevance of code anomalies for identifying architecture degra-
            dation symptoms. In 2012 16Th european conference on software maintenance and
            reengineering, pages 277–286. IEEE, 2012.
            
            \item [55] Giancarlo Sierra, Ahmad Tahmid, Emad Shihab, and Nikolaos Tsantalis. Is self-
            admitted technical debt a good indicator of architectural divergences? In 2019
            IEEE 26th International Conference on Software Analysis, Evolution and Reengi-
            neering (SANER), pages 534–543. IEEE, 2019.
            
            \item [56] Tushar Sharma. How deep is the mud: Fathoming architecture technical debt
            using designite. In 2019 IEEE/ACM International Conference on Technical Debt
            (TechDebt), pages 59–60. IEEE, 2019.
            
            \item [57] Ruiyin Li, Peng Liang, Mohamed Soliman, and Paris Avgeriou. Understanding
            software architecture erosion: A systematic mapping study. Journal of Software:
            Evolution and Process, 34(3):e2423, 2022.
            
            \item [58] Fehmi Jaafar, Salima Hassaine, Yann-Gaël Guéhéneuc, Sylvie Hamel, and Bram
            Adams. On the relationship between program evolution and fault-proneness: An
            empirical study. In 2013 17th European Conference on Software Maintenance and
            Reengineering, pages 15–24. IEEE, 2013.
            
            \item [59] Bernhard Merkle. Stop the software architecture erosion: Building better software
            systems. In Proceedings of the ACM International Conference Companion on Object
            Oriented Programming Systems Languages and Applications Companion, page
            129–138, 2010.
            
            \item [60] Neil A Ernst, Stephany Bellomo, Ipek Ozkaya, Robert L Nord, and Ian Gorton.
            Measure it? manage it? ignore it? software practitioners and technical debt. In
            Proceedings of the 10th Joint Meeting on Foundations of Software Engineering,
            pages 50–60, 2015.
            
            \item [61] Miryung Kim, Thomas Zimmermann, and Nachiappan Nagappan. A field study
            of refactoring challenges and benefits. In Proceedings of the ACM SIGSOFT 20th
            International Symposium on the Foundations of Software Engineering, pages 1–11,
            2012.
            
            \item [62] James Ivers, Robert L. Nord, Ipek Ozkaya, Chris Seifried, Christopher S. Timperley,
            and Marouane Kessentini. Industry experiences with large-scale refactoring. In
            Proceedings of the 30th ACM Joint European Software Engineering Conference and
            Symposium on the Foundations of Software Engineering, ESEC/FSE 2022, page
            1544–1554, 2022.
            
            \item [63] Muhammad Atif Javed and Uwe Zdun. A systematic literature review of trace-
            ability approaches between software architecture and source code. In Proceedings
            of the 18th International Conference on Evaluation and Assessment in Software
            Engineering, pages 1–10, 2014.
            
            \item [64] Mohsen Vakilian, Nicholas Chen, Stas Negara, Balaji Ambresh Rajkumar, Brian P
            Bailey, and Ralph E Johnson. Use, disuse, and misuse of automated refactorings.
            In Proceedings of the 34th International Conference on Software Engineering, pages
            233–243, 2012.
            
            \item [65] Emerson Murphy-Hill, Chris Parnin, and Andrew P Black. How we refactor, and
            how we know it. IEEE Transactions on Software Engineering, 38(1):5–18, 2011.
            \end{itemize}

\subsection{Summary of Challenges with Recommendations}\label{supplement:summary}
The ``Observed in Literature'' column in Table \ref{tab:summary} highlights the literature that discuss the corresponding challenges. 
\begin{table*}[t]
      \centering
      \scriptsize
    \caption{Summary of challenges with recommendations.}
    \vspace{-0.4cm}
      \begin{tabular}{p{5.8cm}p{1.2cm}p{46em}}
      \toprule
      \textbf{Challenge Description } & \textbf{Observed in Literature} & \multicolumn{1}{p{46em}}{\textbf{Recommendations}} \\
      \hline
      \rowcolor[rgb]{ .906,  .902,  .902} \textbf{SOFTWARE REQUIREMENTS} &   &  \\
      Unpredictable evolution and changes of software requirements complicate architecture design (\faGroup,~\faCalendar). \vspace{0.6cm} & \cite{dasanayake2019impact,szlenk2012modelling,de2012controlling,mens2008software,tang2011software,arcelli2015control,baudry2014diversify}      & \faCalendar~Make tradeoffs among multiple factors while being aware of the volatility and unpredictability of requirements. 
    \newline{}\faFileText~Use formal architecture documentation to capture design tradeoffs, and establish trace links between requirements and design decisions. 
    \newline{}\faCalendar~Establish a standardized process for adapting, refactoring, and retiring software architectures. \\
    \rowcolor[rgb]{ .906,  .902,  .902} \textbf{SOFTWARE DESIGN} &       &  \\
    \rowcolor[rgb]{ .906,  .902,  .902} \textit{\textbf{$\blacktriangleright$ Design Documentation}} &       &  \\
    Use of models and tools are inadequate to ensure the completeness of architecture documentation (\faFileText,~\faCogs).& \cite{zhi2015cost}      & \multirow{3}[0]{46em}{\faCalendar~Implement a standardized process where developers update design documentation before merging code, and utilize wikis for hosting the documentation.\newline{}\faCogs~Utilize collaborative writing tools to generate and share design documentation.\newline{}\faCalendar~Apply the ``code as documentation'' principle to minimize costs and efforts in writing and maintaining design documentation.} \\
    Architecture documentation becomes obsolete as software evolves (\faFileText,~\faCalendar). &   \cite{brown2018c4}  &  \\
    Inadequate tool support for sharing, version control, and tracing of scattered design documentation (\faFileText,~\faCogs).&   \cite{capilla2016ten}   &  \\
    \rowcolor[rgb]{ .906,  .902,  .902} \textit{\textbf{$\blacktriangleright$ Design Principles}} &       &  \\
    Unclear boundaries between architectural elements in software systems (\faGroup,~\faCalendar). &  \cite{lung2004applications,mitchell2006automatic,li2019dataflow,chen2017monolith}     & \multirow{2}[0]{46em}{\faCalendar~Isolate software from physical hardware changes in software systems with layered architecture styles. \newline{}\faCalendar~Take into account the underlying business requirements while decomposing cloud-based and microservice systems into microservices and components. \newline{}\faCalendar~Apply Domain-Driven Design for flexible and iterative microservices architecture design.} \\
      Interdisciplinary knowledge is required to lower coupling and improve cohesion of software (\faGroup,~\faCalendar). \vspace{0.1cm}&  --   &  \\

      \rowcolor[rgb]{ .906,  .902,  .902} \textit{\textbf{$\blacktriangleright$ Design Quality Analysis and Evaluation}} &       &  \\
      Architecture review requires a standard process, active involvement of external experts, and tool support (\faCalendar,~\faGroup,~\faCogs). &   \cite{babar2009software}  &\multirow{4}[0]{46em}{\faGroup~Engage external experts in architecture review and evaluation, and offering prioritized recommendations for architecture improvement. \newline{}\faCalendar~Employ simulation and prototyping to evaluate solutions for architecture decisions.\newline{}\faCogs~Automate a limited set of measures to quantify architecture quality within organizations, leaving room for practitioners to make decisions. \newline{}\faCogs~Employ whitelists to exclude special cases when measuring architecture quality.} \\
      Lack of effective and apply-to-all quantitative measures (\faCogs,~\faCalendar).\vspace{0.9cm} &  --  &  \\
      \rowcolor[rgb]{ .906,  .902,  .902} \textbf{SOFTWARE CONSTRUCTION AND TESTING} &       &  \\
      \rowcolor[rgb]{ .906,  .902,  .902} \textit{\textbf{$\blacktriangleright$ Architecture Conformance Checking}} &       &  \\
      Automated architecture conformance checking is rare (\faCogs). &   \cite{caracciolo2014software,thomas2017static}  & \multicolumn{1}{l}{\multirow{2}[0]{46em}{\faGroup~Avoid knowledge vaporization to maintain consistency between the intended architecture and implemented system.}} \\
      Obsolete documentation and lack of traceability hinder automation of architecture conformance checking (\faFileText,~\faCalendar). &   --  &  \\
      \rowcolor[rgb]{ .906,  .902,  .902} \textit{\textbf{$\blacktriangleright$ Continuous Architecture Monitoring}} &       &  \\
      Limited tool support to continuously monitor the health status of software architectures (\faCogs). &   \cite{soares2023trends}  & \multicolumn{1}{l}{\multirow{3}[0]{46em}{\faCalendar~Leverage automated testing frameworks to enable continuous monitoring of system performance. \newline{}\faCogs~Develop tools that facilitate continuous architecture monitoring. \newline{}\faCalendar~Allocate resources for maintaining architecture monitoring tools, or establish a dedicated maintenance team. \newline{}\faGroup~Involve architects in the identification of architecture problems during continuous architecture monitoring.}} \\
      Pinpointing architecture problems requires a system-wide perspective (\faCalendar). &  \cite{wuttke2010automatically}   &  \\
      Maintenance of monitoring tools requires ongoing effort (\faGroup). \vspace{0.1cm} &   --  &  \\
    
      \rowcolor[rgb]{ .906,  .902,  .902} \textit{\textbf{$\blacktriangleright$ Construction Quality}} &       &  \\
      Technical debts are introduced to software projects (\faGroup,~\faCalendar). &   \cite{brunet2012evolutionary}  & \multicolumn{1}{l}{\multirow{3}[0]{46em}{\faGroup, \faCalendar~Invest effort and time in addressing technical debts. \newline{}\faCogs, \faCalendar~Automate architectural smell detection to streamline the enforcement of architectural constraints. 
      \newline{} \faGroup, \faCogs~Promote collaboration among project teams to identify organizational-wide dependency-related architectural smells.}} \\
      Lack of tool support for detecting architectural smells (\faCogs). &  \cite{mumtaz2021systematic}   &  \\
      Unawareness of correlation between code smells and architecture problems (\faGroup). \vspace{0.1cm} &  \cite{li2011case,guimaraes2015architecture,macia2012automatically,bertran2011detecting}   &  \\
      \rowcolor[rgb]{ .906,  .902,  .902} \textbf{SOFTWARE MAINTENANCE} &       &  \\
      \rowcolor[rgb]{ .906,  .902,  .902} \textit{\textbf{$\blacktriangleright$ Architecture Erosion}} &       &  \\
      Lack of tool support to explicitly capture and aggregate symptoms of architecture erosion (\faCogs). &   --  & \multicolumn{1}{l}{\multirow{2}[0]{46em}{\faCogs~Construct a dashboard for visualizing architecture erosion symptoms gathered from various sources.}} \\
      Obsolete documentation and increasing complexity of software systems accelerate architecture erosion (\faCalendar,~\faGroup). &   \cite{jaafar2013relationship,de2012controlling,merkle2010stop,macia2012relevance}  &  \\
      \rowcolor[rgb]{ .906,  .902,  .902} \textit{\textbf{$\blacktriangleright$ Architecture Refactoring}} &       &  \\
      No agreement on the value of architecture refactoring (\faGroup). &  \cite{ernst2015measure,kim2012field}   & \multicolumn{1}{l}{\multirow{3}[1]{46em}{\faGroup~Foster a software engineering culture where the entire organization collectively prioritizes high-quality software architecture. \newline{}\faCalendar~Implement a structured process to establish traceability between design and implementation artifacts.}} \\
      Inadequate tool support for impact analysis of architectural changes (\faCogs). &   \cite{kim2014empirical,sharma2015challenges}  &  \\
      Inadequate tool support for module- and system-level refactoring (\faCogs). &  \cite{vakilian2012use,murphy-hill2011we,ivers2022industry}
      &  \\
      \bottomrule
    \end{tabular}\label{tab:summary}\vspace{-0.4cm}
    \end{table*} 

\end{document}